\documentclass[a4paper,11pt]{article}

\usepackage{jheppub} 
\usepackage[compat=1.0.0]{tikz-feynman}
\usepackage{graphicx,subfig}
\usepackage{physics}
\usepackage{amsmath}
\usepackage{tikz}
\usepackage[font=footnotesize,labelfont=bf]{caption}
\usepackage{appendix}
\usepackage{amssymb}
\usepackage[flushleft]{threeparttable}

\usetikzlibrary{external}
\immediate\write18{mkdir -p pgf-img}
\newcommand{\intl}[2]{\int\limits_{#1}^{#2}}
\newcommand{\bff}[1]{\mathbf{#1}}
\newcommand{\dagg}{^\dagger}

\newcommand{\intsp}{\int d^4}
\newcommand{\intmo}[1]{\int\frac{d^4{#1}}{(2\pi)^4}~}

\subheader{\begin{flushright}
		\texttt{UTTG 24-2021}
\end{flushright}}

\title{Quantum chaos in a weakly-coupled field theory with nonlocality}

\author[a]{Willy Fischler}
\author[a]{,Tyler Guglielmo}
\author[b,c]{, and Phuc Nguyen}

\affiliation[a]{Theory Group, Department of Physics, The University of Texas at Austin,\\Austin, Texas 78712, USA}
\affiliation[b]{Department of Physics and Astronomy, Lehman College, City University of New York,\\
250 Bedford Park Blvd. W, Bronx NY 10468, USA}
\affiliation[c]{Department of Mathematics and Haifa Research Center for Theoretical Physics and Astrophysics, University of Haifa,\\
Haifa 3498838, Israel}

\emailAdd{fischler@physics.utexas.edu}
\emailAdd{tylerg@utexas.edu}
\emailAdd{phuc.nguyen@lehman.cuny.edu}

\abstract{In order to study the chaotic behavior of a system with non-local interactions, we will consider weakly coupled non-commutative field theories.  We compute the Lyapunov exponent of this exponential growth in the large Moyal-scale limit to leading order in the t'Hooft coupling and $1/N$. We found that in this limit, the Lyapunov exponent remains comparable in magnitude to (and somewhat smaller than) the exponent in the commutative case. This can possibly be explained by the infrared sensitivity of the Lyapunov exponent. Another possible explanation is that in examples of weakly coupled non-commutative field theories, non-local contributions to various thermodynamic quantities are sub-dominant.}

\begin{document} 
\maketitle
\flushbottom

\section{Introduction}
Since it was proposed by Kitaev that out-of-time-order correlation functions (OTOCs) are a useful measure of the butterfly effect \cite{KitaevTalk}, OTOCs and quantum chaos in general have become an important aspect of AdS/CFT and holography \cite{Shenker:2013pqa, Shenker:2014cwa, Maldacena:2015waa, Maldacena:2016hyu, Maldacena:2017axo}.  Inspired by the work of Douglas Stanford, we consider a non-commutative extension of his matrix model, and along the same vein, a vector model \cite{Stanford:2015owe}.  For a selection of recent work on the topic of chaos and holography, see \cite{deBoer:2017xdk, Fischler:2018kwt, Couch:2019zni, Eccles:2021zum, Romero-Bermudez:2019vej, Liao:2018uxa, Kundu:2021qcx,Kundu:2021mex}. The most important property of OTOCs (for the purpose of chaos) is that these correlators exhibit a regime of exponential growth in time, from which we can identify a Lyapunov exponent.

In this work, we extend existing studies of OTOCs to non-commutative field theories. These are examples of non-local field theories, and are obtained from the more familiar field theories such as $\lambda \phi^{4}$ by replacing the usual multiplication of fields by the so-called Moyal product. For a selection of work on non-commutative field theory and holography, see \cite{Edalati:2012jj,Couch:2017yil}. Specifically, we will consider two examples of non-commutative field theories: a vector $O(N)$ model, and a theory with hermitian matrices; in both cases, we will compute the Lyapunov exponent in the t'Hooft limit.

The action of the non-commutative vector $O(N)$ model is:
\begin{equation}
S_{vector} = \intsp{x} \left( \frac12 \partial_{\mu}\phi_{i} \star \partial^{\mu}\phi_{i} - \frac12 m^2\phi_{i} \star \phi_{i} - \frac{\lambda_{1}}{4 N}\phi_{i} \star \phi_{i} \star \phi_{j} \star \phi_{j} - \frac{\lambda_{2}}{4 N}\phi_{i} \star \phi_{j} \star \phi_{i} \star \phi_{j} \right)
\end{equation}
where the $\phi_{i}$'s are real scalar fields, and the $*$-product (also known as the Moyal product) is defined by
\begin{equation}
    (f \star g)(x) = E(x,y) f(x) g(y)|_{x=y}
\end{equation}
where $E(x,y) \equiv e^{\frac{i}{2}\Theta^{\mu\nu}\partial_{x^\mu}\partial_{y^\nu}}$ and $\Theta^{\mu\nu}$ an anti-symmetric matrix that characterizes the non-commutativity. Even though the Moyal product between two functions is in general a complex quantity, one can check that the action above is real. To see this, first note that, after an integration by parts, the free-field part of the action above can be checked to be insensitive to the value of $\theta$.  Therefore it is the same as the commutative free-field theory.  As for the interaction terms, it can be seen that $\phi \star \phi (x)$ is real for any real-valued function $\phi$, from which it follows that both interaction terms above are real \footnote{To see that $\phi \star \phi (x)$ is real, we can expand the exponential in the definition of the Moyal product, and argue that the terms containing an odd number of $\theta^{\mu\nu}$ vanish by symmetry/antisymmetry of the indices. Alternatively, we can use the fact that $(f \star g)^{*} = g^{*} \star f^{*}$ for any two complex-valued functions $f$ and $g$. In other words, complex conjugation is an antilinear antiautomorphism.}.

The action of the hermitian matrix model is:
\begin{equation}
   S_{matrix} = \int d^{4}x \left[ \frac{1}{2} \mathrm{Tr}({\partial_{\mu} \Phi \partial^{\mu} \Phi}) - \frac{1}{2}m^{2} \mathrm{Tr}{(\Phi^{2})} - \frac{\lambda_1}{4N} \Phi^a_b \star \Phi^b_c \star \Phi^c_d \star \Phi^d_a - \frac{\lambda_2}{4N}\Phi^a_b \star \Phi^b_c \star \Phi^d_a \star \Phi^c_d \right]
\end{equation}
In the above action, $\Phi_{ab}$ is a hermitian matrix, and the last two terms can be shown to be real as in the vector model.

The Lyapunov exponent will be extracted from the following ``commutator-squared'' OTOC in the vector case:
\begin{equation} \label{eq:OTOC}
    C_{\text{vector}}(T) \equiv -\frac{1}{N^2} \int d^{3} \textbf{x}~ \mathrm{Tr}{\left(\sqrt{\rho} [\phi_{i}(T,\textbf{x}), \phi_{j}(t_{0},0)] \sqrt{\rho} [\phi_{i}(T,\textbf{x}), \phi_{j}(t_{0},0)] \right)}
\end{equation}
where $\rho = \frac1Z e^{-\beta H}$ is the thermal density matrix. The above can also be written as the following thermal expectation value:
\begin{equation}
    C_{\text{vector}}(T) \equiv -\frac{1}{N^2} \int d^{3} \textbf{x}~ \langle [ \phi_{i}(T-i\beta/2,\textbf{x}),\phi_{j}(t_{0}-i\beta/2,0) ] [ \phi_{i}{(T,\textbf{x})},\phi_{j}(t_{0},0) ]  \rangle_{\beta}
\end{equation}
From the second form of $C(T)$, we see in particular that the four operators are equally spaced around the thermal circle.

The matrix model's Lyapunov exponent will be extracted from a similar OTOC given by \cite{Stanford:2015owe},
\begin{align}
	C_{\text{matrix}}(T) &\equiv \frac{1}{N^4} \sum_{aba'b'} \int d^3 \bff{x} ~\Tr(\sqrt{\rho} \comm{\Phi_{ab}(t,\bff{x})}{\Phi_{a'b'}} \sqrt{\rho} \comm{\Phi_{ab}(t,\bff{x})}{\Phi_{a'b'}} \dagg) 
\end{align}

Our main motivation is to see how the Lyapunov exponent is affected when we turn on a Moyal scale, i.e. the scale characterizing the non-commutativity or non-locality of the field theory. Naively, we expect that non-locality enhances chaos, so the Lyapunov exponent should increase with the Moyal scale.  However, what we find is that in the limit of large Moyal scale, keeping all the other parameters fixed, the Lyapunov exponent's dependence on the various parameters of the theory does not change.

On a technical level, we follow the methodology of \cite{Stanford:2015owe}, where the Lyapunov exponent of a weakly-coupled matrix theory was derived using the Schwinger-Keldysh formalism.

The rest of the paper is organized as follows. In section \ref{sec:Prelim}, we review the Schwinger-Keldysh formalism as a means to compute OTOCs, and also review a few basic facts about non-commutative field theories. In section \ref{sec:LyapunovVector}, we compute the Lyapunov exponent in the vector model case. In section \ref{sec:LyapunovMatrix}, we compute the Lyapunov exponent in the matrix case. In section \ref{sec:Conclusion}, we discuss the implications of our findings and conclude. We explain a few technical points in the appendices.

\section{A few preliminaries}\label{sec:Prelim}
In this section, we explain the Feynman rules for the computation of a ``commutator-squared'' OTOC with non-commutativity. For simplicity, in this section we will suppress the group index of the scalar field, and think about non-commutative $\lambda \phi^{4}$ theory with a single scalar field. The $O(N)$ group structure does not play an important role in this section.

\subsection{Vertex factor}
First, let's just consider the computation of an ordinary time-ordered correlation function of a non-commutative theory. As derived in \cite{Minwalla:1999px}, the only modification to the Feynman rules due to non-commutativity is in the form of a phase factor associated to each interaction vertex. To derive that factor, first consider the Moyal product of four scalar fields:
\begin{align}
\phi\star\phi\star\phi\star\phi(x) &= E(x,y)E(x,z)E(x,w)E(y,z)E(y,w)E(z,w) \phi(x)\phi(y)\phi(z)\phi(w)|_{x=y=z=w}.
\end{align}
In momentum space, the interaction part of the action then takes the form:
\begin{align}
S_{\text{int}} &= (2\pi)^4 \intmo{k_1}\intmo{k_2}\intmo{k_3}\intmo{k_4} \delta(k_1 + k_2 + k_3 + k_4) V(k_1, k_2, k_3, k_4) \times \nonumber\\
&\times \phi_{k_1}\phi_{k_2}\phi_{k_3}\phi_{k_4} \label{eq:fTrans}
\end{align}
where we define
\begin{align}
V(k_1,\dots, k_n) \equiv e^{-\frac{i}{2}\sum\limits_{i<j} \Theta \cdot k_{i} \cdot k_{j}}
\end{align}
where $\Theta \cdot k_{i} \cdot k_{j} \equiv \Theta^{\mu\nu} k_{i,\mu} k_{j,\nu}$. The above is the vertex factor of \cite{Minwalla:1999px}. Graphically speaking, we have the Feynman rule
\begin{equation}
		\feynmandiagram [baseline=(a.base), horizontal = i1 to f2] {
			i1 -- [momentum={\(k\)}] a[dot]  -- [rmomentum=\(k\)] f1,
			i2 -- [momentum=\(k\)] a  -- [rmomentum=\(k\)] f2,
		}; 
		~= \frac{-i\lambda}{4!} (\text{\# of Wick Contractions}) (2\pi)^4 \delta(k_1 + k_2 + k_3 + k_4) V(k_1, k_2, k_3, k_4) \nonumber
	\end{equation}
It can be shown that $V$ is invariant under a cyclic permutation of the four momenta $k_{1}$, $\dots$, $k_{4}$. Thus, we have to keep track of the cyclic ordering of the momenta around each vertex in a given Feynman diagram. As a result, planar contractions are in general not equivalent to non-planar contractions, because no cyclic permutation takes a planar contraction to a non-planar one.

\subsection{Schwinger-Keldysh formalism}
Next, we review how to use the Schwinger-Keldysh formalism to compute the following OTOC (without commutators):
\begin{eqnarray}
    \tilde{C}(T) &\equiv& -\langle  \phi(T-i\beta/2) \phi(t_{0}-i\beta/2) \phi{(T)} \phi(t_{0}) \rangle_{\beta}
\end{eqnarray}
for the non-commutative $\lambda \phi^{4}$ theory. To ease notation, we have suppressed the spatial locations of the four operators.

The first step is to express the Heisenberg operators above in terms of interaction-picture operators $\phi^{I}$, by the relation $\phi(t) =S^{\dagger}{(t,t_{0})} \phi^{I}{(t)} S{(t,t_{0})}$, with $S(t,t_{0}) \equiv e^{iH_{0}(t-t_{0})}e^{-iH(t-t_{0})}$. For simplicity, we have chosen the reference time at which the Heisenberg picture, interaction picture and Schrodinger picture agree to be the same as the time $t_{0}$ at which two of the four operators are inserted. We will keep $t_{0}$ general in this subsection, and set it to $0$ afterward.

Also, we can write $e^{-\beta H}$ as $e^{-\beta H_{0}} S(t_{0}-i\beta,t_{0})$. Then:
\begin{eqnarray}\label{Ctilde}
\tilde{C}{(T)} &=& -\frac{1}{Z} \mathrm{Tr} \bigg [ e^{-\beta H_{0}} S(t_{0}-i\beta, T-i\beta/2) \phi^{I}{(T-i\beta/2)} S(T-i\beta/2, t_{0}-i\beta/2) \nonumber \\
&& \phi^{I}{(t_{0}-i\beta/2)} S(t_{0}-i\beta/2,T) \phi^{I}{(T)} S(T,t_{0}) \phi^{I}{(t_{0})} \bigg ]
\end{eqnarray}
It is useful to visualize the four interaction-picture operators as inserted at four locations along a contour in the complex-time plane, as depicted in Fig. \ref{fig:SKContour} below. 
\begin{figure}[h]
\centering
\includegraphics[width=0.3\textwidth]{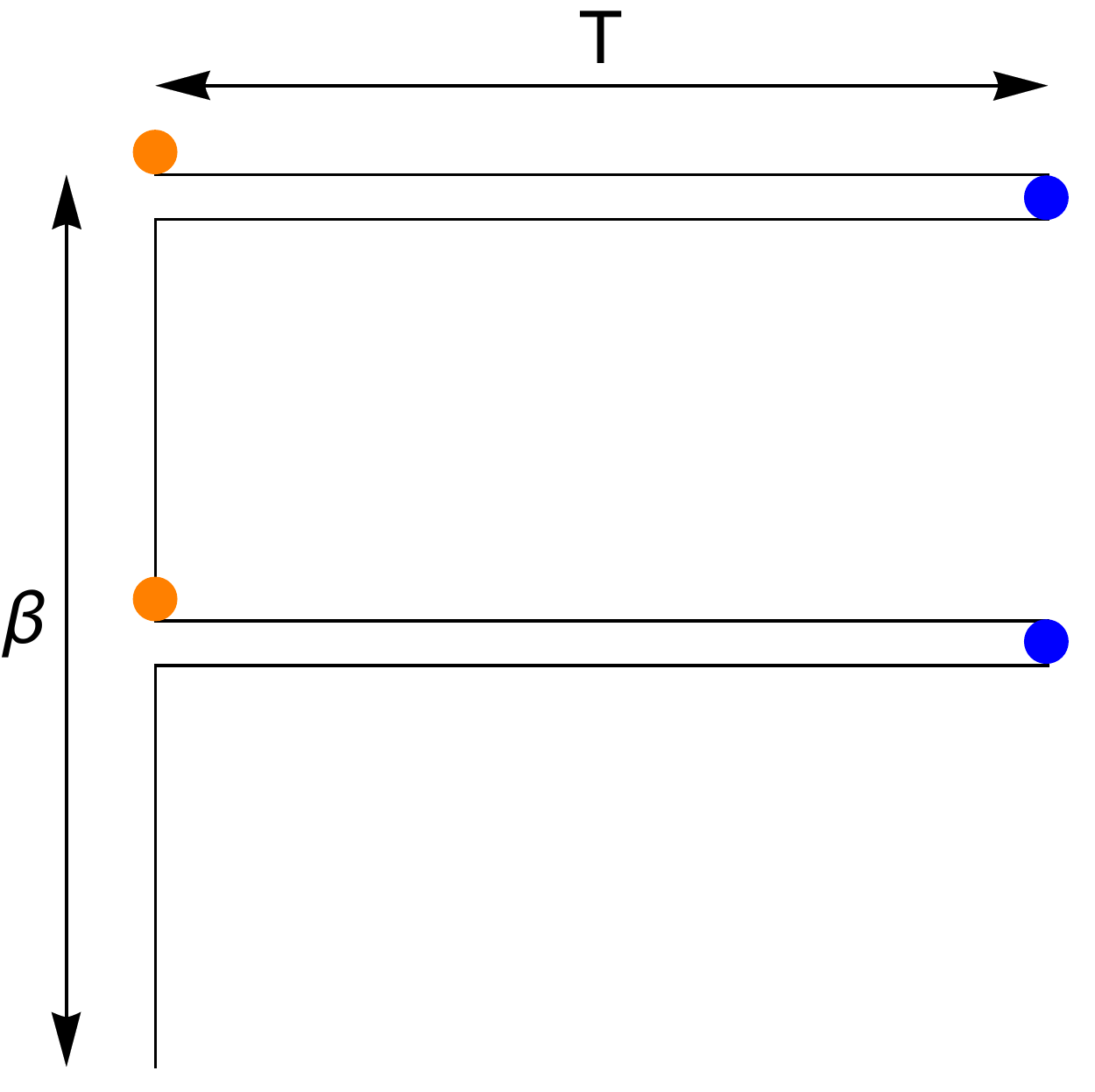}
\caption{Schwinger-Keldysh contour for $\tilde{C}{(T)}$ for the special case $t_{0} = 0$.}
\label{fig:SKContour}
\end{figure}
We will refer to this contour as the Schwinger-Keldysh (SK) contour.  We note that the two timefolds are separated by half the thermal circle -- this is the usual setup in the literature on OTOCs, although less symmetrical choices of the Schwinger-Keldysh contour have also been studied \cite{Romero-Bermudez:2019vej,  Liao:2018uxa}. Throughout this paper, we stick to the standard ``symmetrical'' Schwinger-Keldysh contour drawn in Fig. \ref{fig:SKContour}.

The usefulness of the SK contour comes from the fact that the operators appearing on the right-hand side of (\ref{Ctilde}) are contour-ordered. Indeed, the four $\phi^{I}$'s appear in (\ref{Ctilde}) in the same order in which they appear along the contour. Furthermore, each of the $S$-operators is itself given by a Dyson series, which is also contour-ordered. So, we can rewrite (\ref{Ctilde}) in the form:
\begin{equation}\label{Ctilde2}
    \tilde{C}{(T)} = -\frac{1}{Z} \mathrm{Tr}{\left[ e^{-\beta H_{0}} T_{\mathcal{C}} \left( e^{-i\int_{\mathcal{C}} dc H_{int}{(c)}} \phi^{I}{(T-i\beta/2)} \phi^{I}{(t_{0}-i\beta/2)} \phi^{I}{(T)} \phi^{I}{(t_{0})} \right) \right]}
\end{equation}
where $T_{\mathcal{C}}$ is contour-ordering, $c$ is the time which elapses along the contour, and $H_{int}$ is the interaction Hamiltonian written in the interaction picture. On the right-hand side of (\ref{Ctilde2}), we now note that what we have is a free-field-theory thermal expectation value. Therefore, the form (\ref{Ctilde2}) is suitable for doing perturbation theory using the standard techniques of Wick's theorem and Feynman diagrams.

Next, we move on to discuss the commutator-squared OTOC:
\begin{eqnarray}
    \hat{C}(T) &\equiv& -\langle  [\phi(T-i\beta/2), \phi(t_{0}-i\beta/2)] [\phi{(T)}, \phi(t_{0})] \rangle_{\beta}
\end{eqnarray}
After expanding both commutators, we see that $\hat{C}$ is a sum of four terms. Now, let's consider the following 1-loop contraction's contribution to $\hat{C}{(T)}$:
\begin{center}
	\begin{tikzpicture}
		\begin{feynman}
			\vertex (i1) at (0,0);
			\vertex (i2) at (4,0);
			\vertex (f1) at (0,2);
			\node[dot] (looppoint) at (2,2);
			\vertex (f2) at (4,2);
		
			\diagram* [] {
				(f1) -- (looppoint) -- (looppoint) -- (f2),
				(i1) -- (i2),
			};
			
			\path (looppoint) --++ (-90:0.5) coordinate (A);
			\draw (A) circle(0.5);
		\end{feynman}
	\end{tikzpicture}
\end{center}
For each of the 4 terms making up $\hat{C}$, the interaction vertex in the diagram above can be on either the upper or lower half of the upper time-fold. Hence, the 4 terms can be split into 8 terms, as depicted in Fig. \ref{8Configurations} below.

By adding up all 8 configurations, we can check that all four horizontal propagators in the diagram above are replaced by a retarded propagator (as opposed to a Wightman function). 

\begin{center}
	\begin{figure}[h!]
		\subfloat[]{\includegraphics[width = 1.5in]{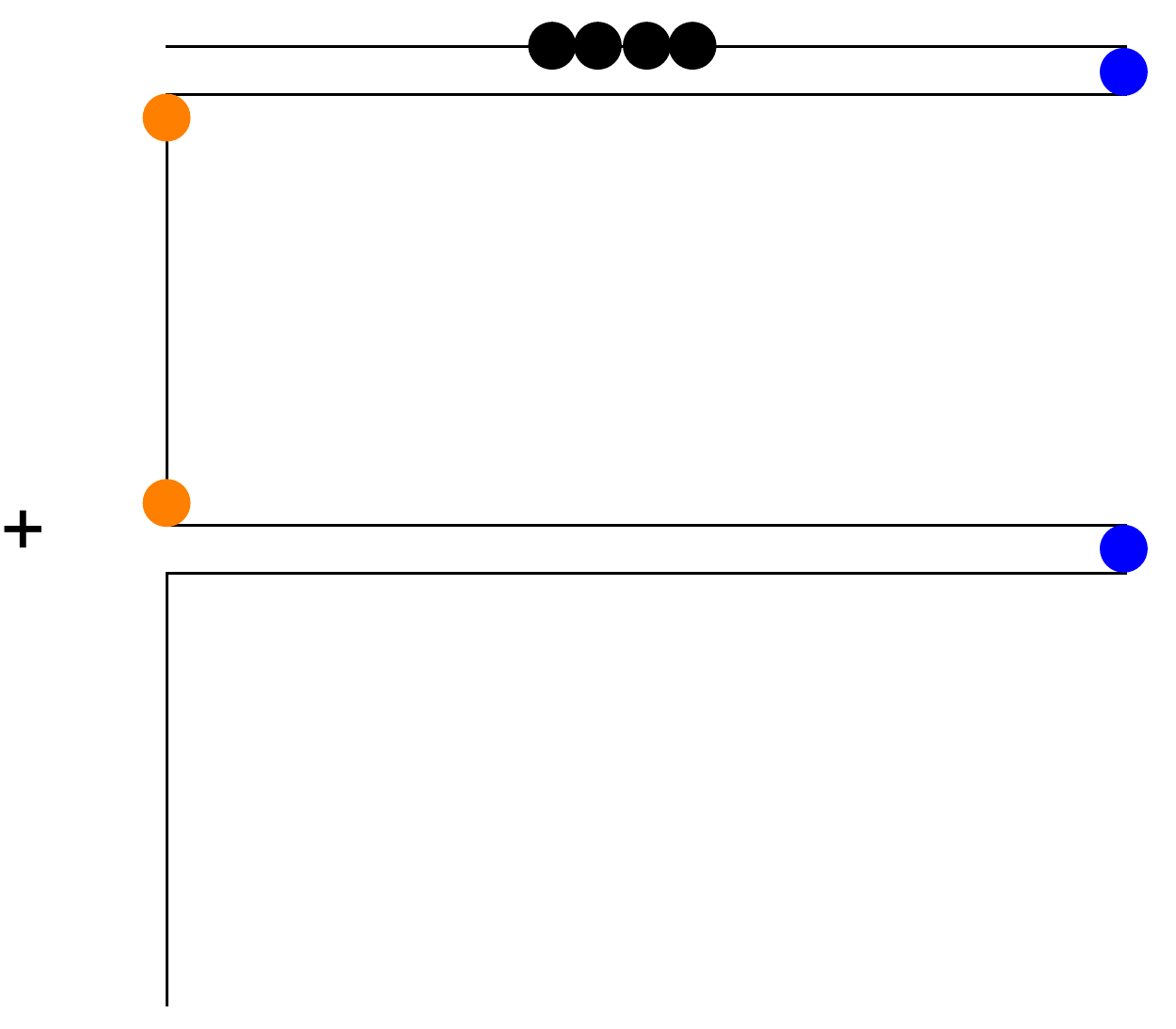}} 
		\subfloat[]{\includegraphics[width = 1.5in]{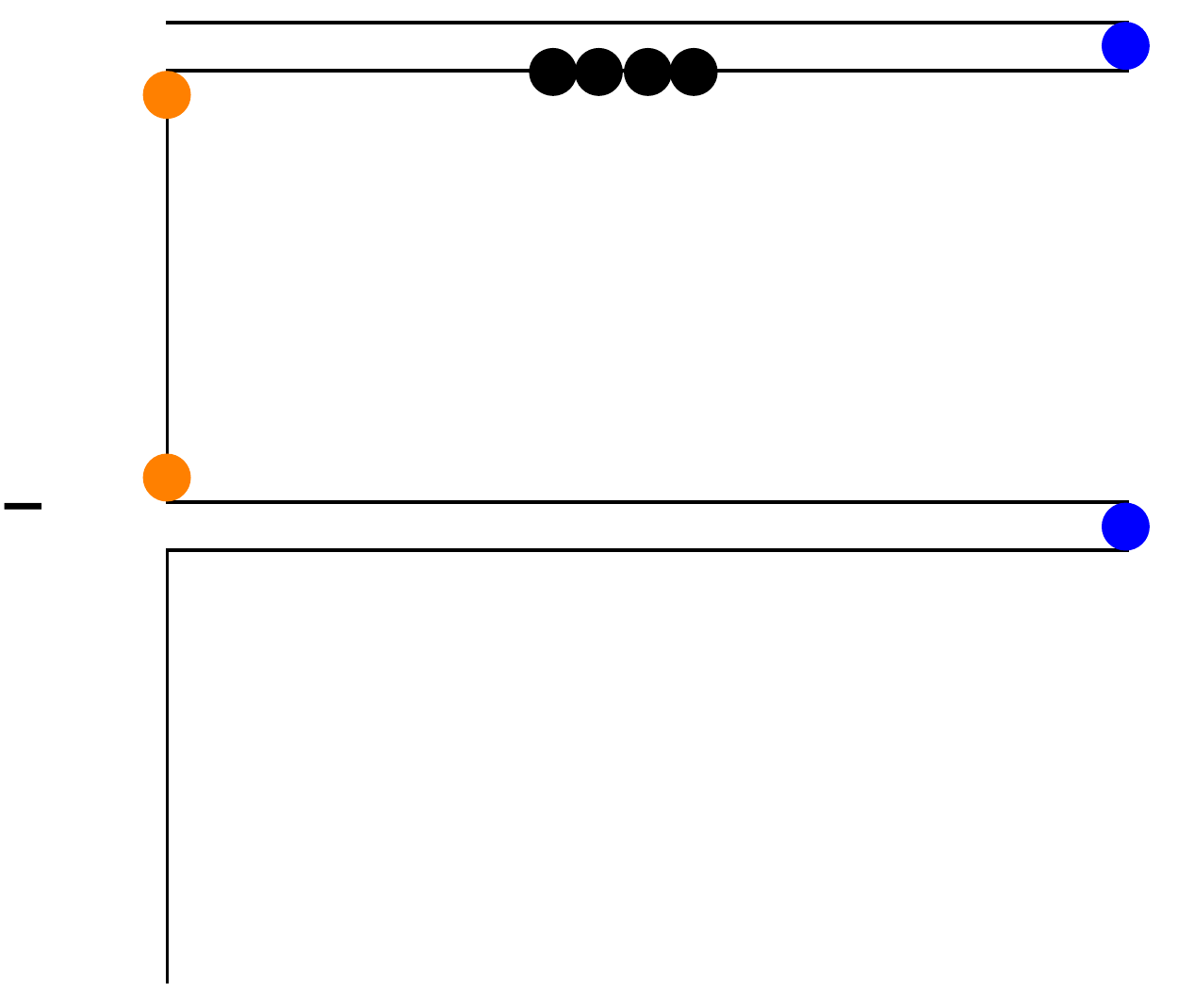}}
		\subfloat[]{\includegraphics[width = 1.5in]{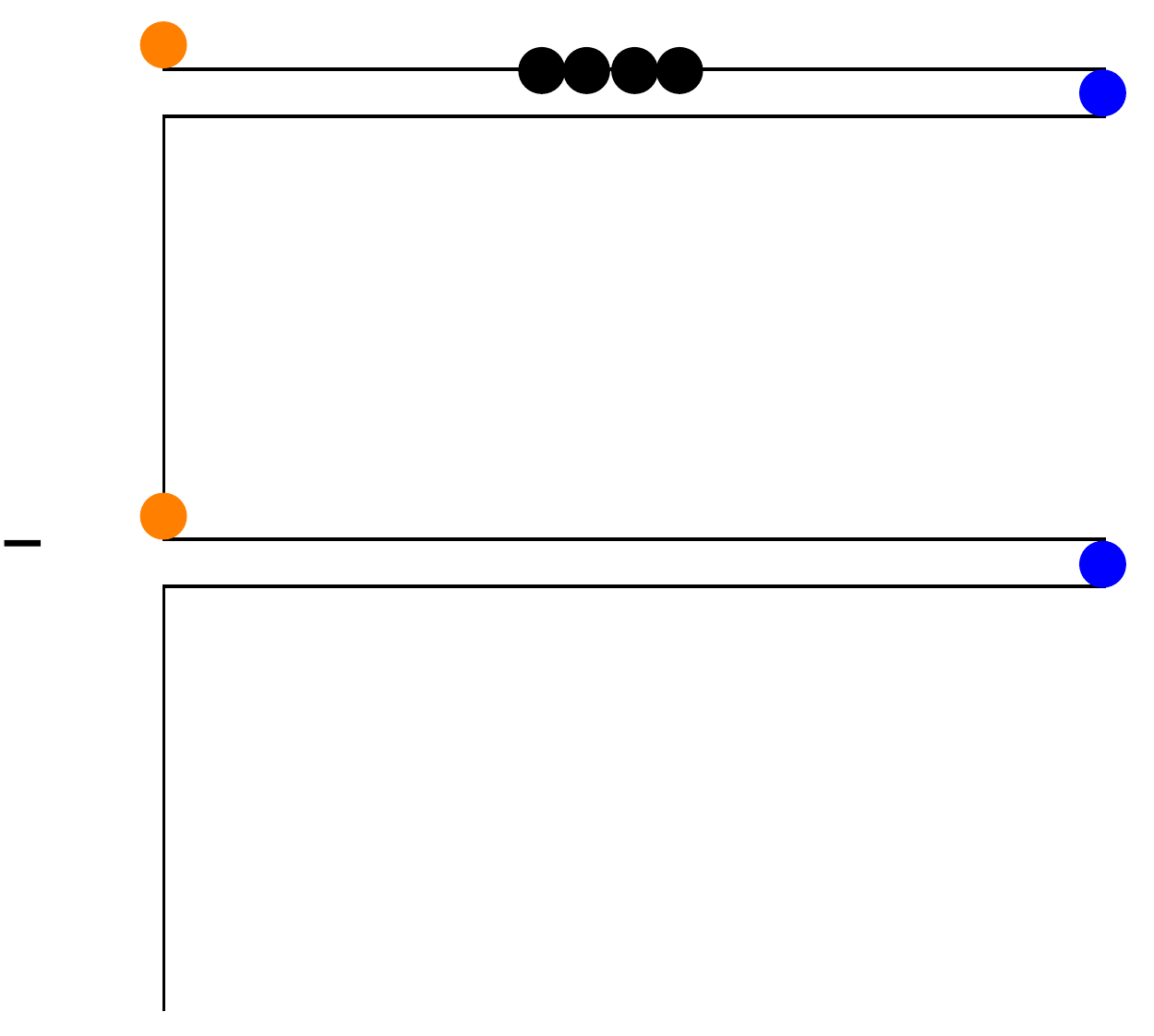}}
		\subfloat[]{\includegraphics[width = 1.5in]{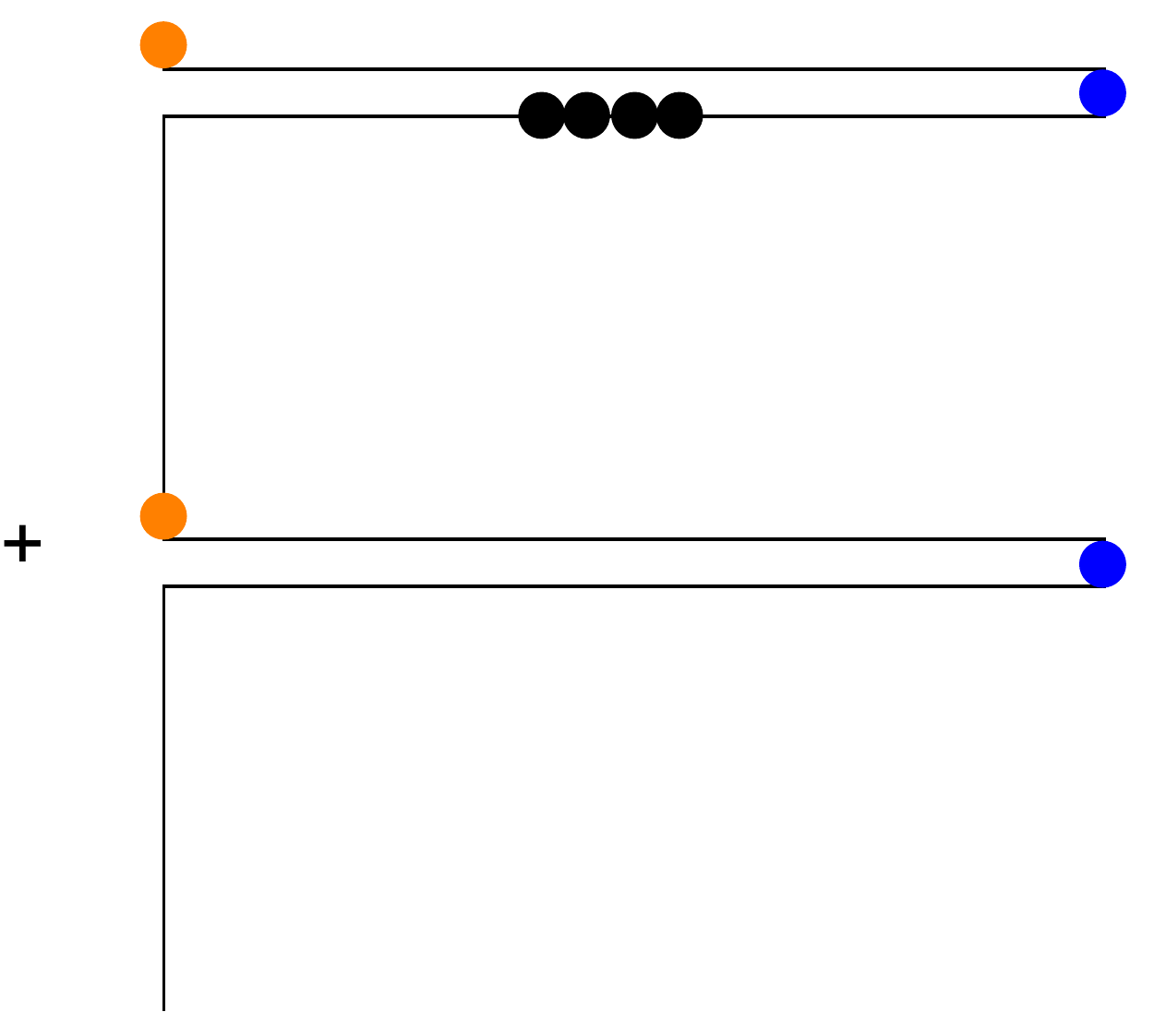}}\\
		\subfloat[]{\includegraphics[width = 1.5in]{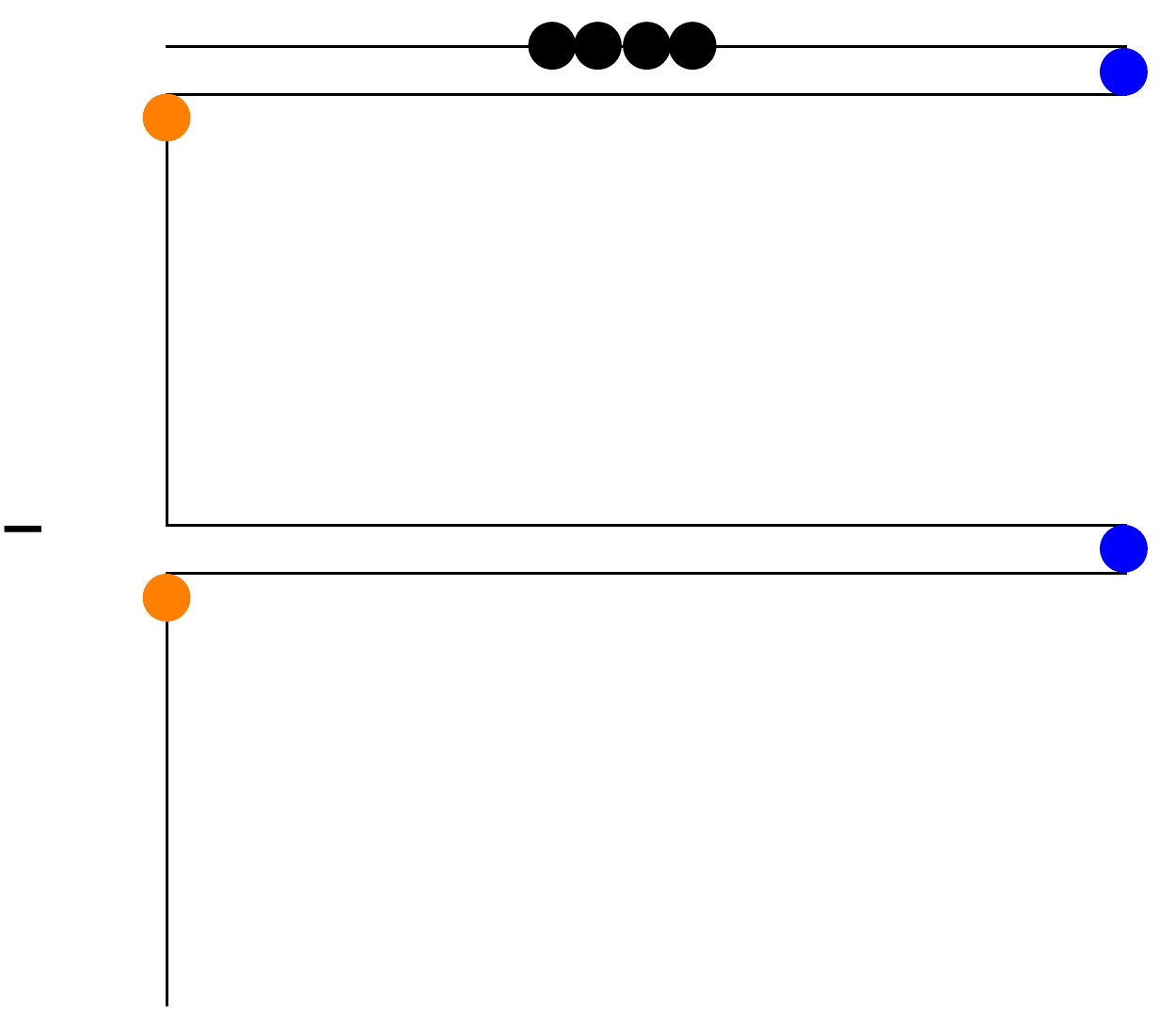}} 
		\subfloat[]{\includegraphics[width = 1.5in]{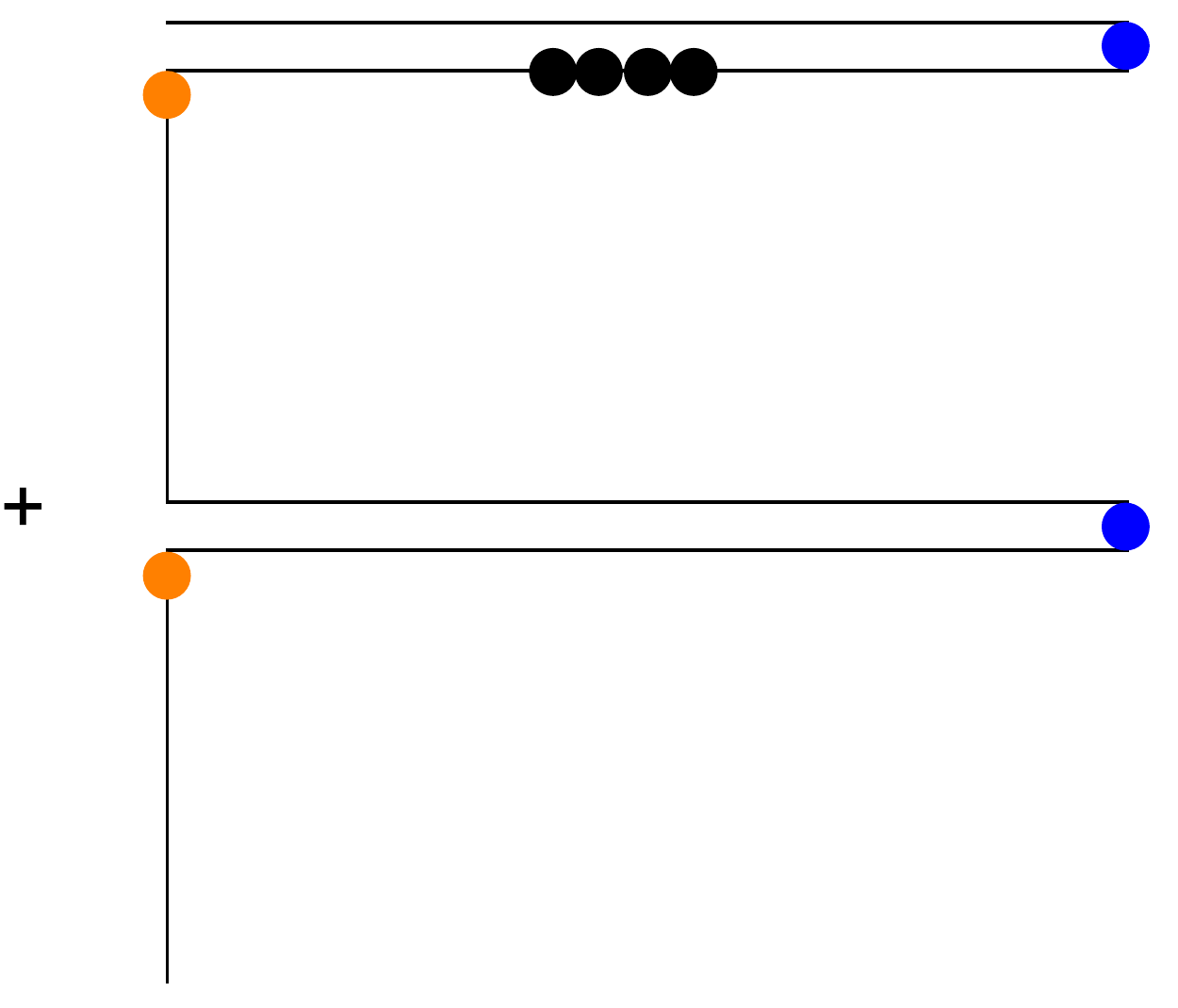}}
		\subfloat[]{\includegraphics[width = 1.5in]{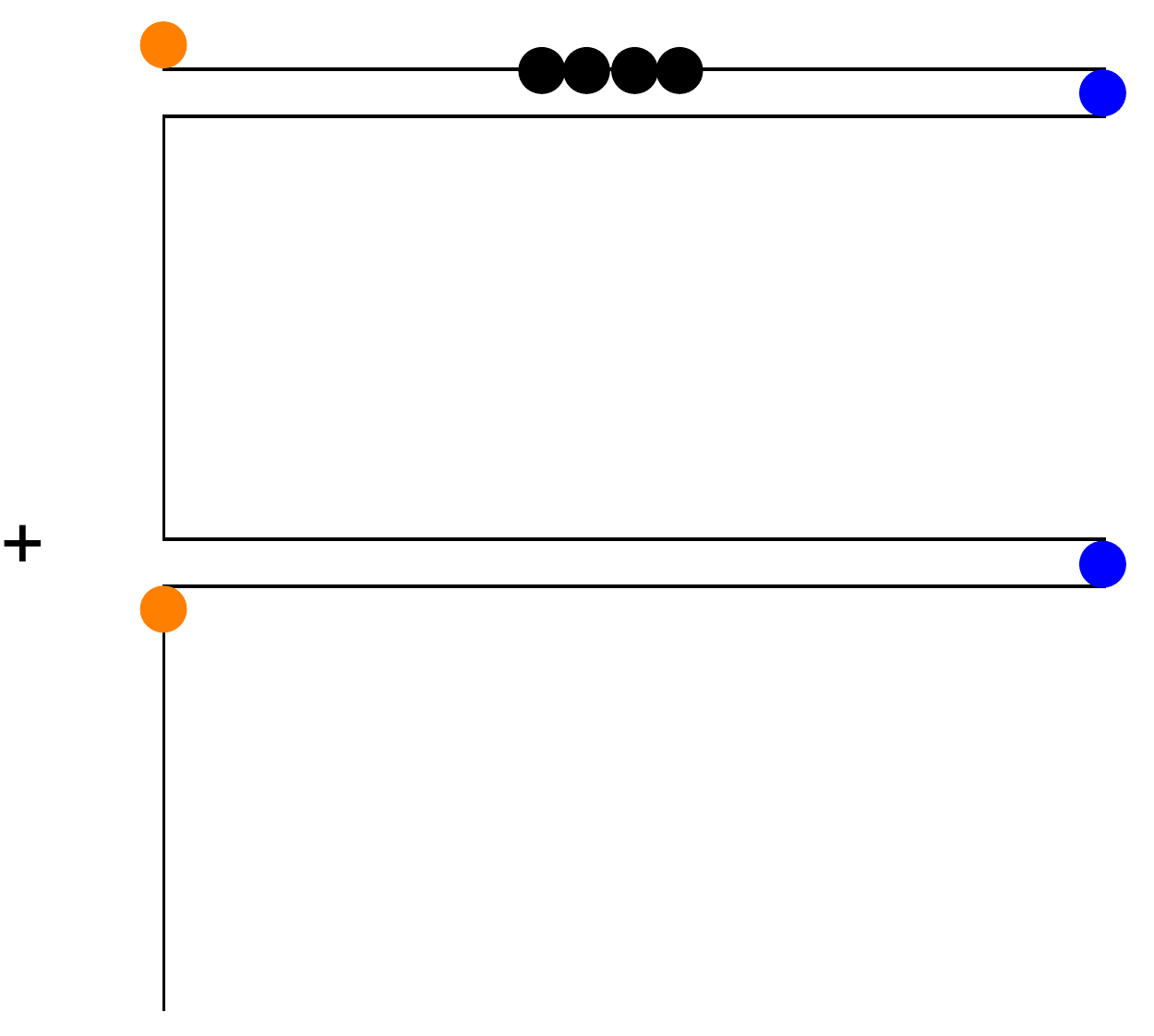}}
		\subfloat[]{\includegraphics[width = 1.5in]{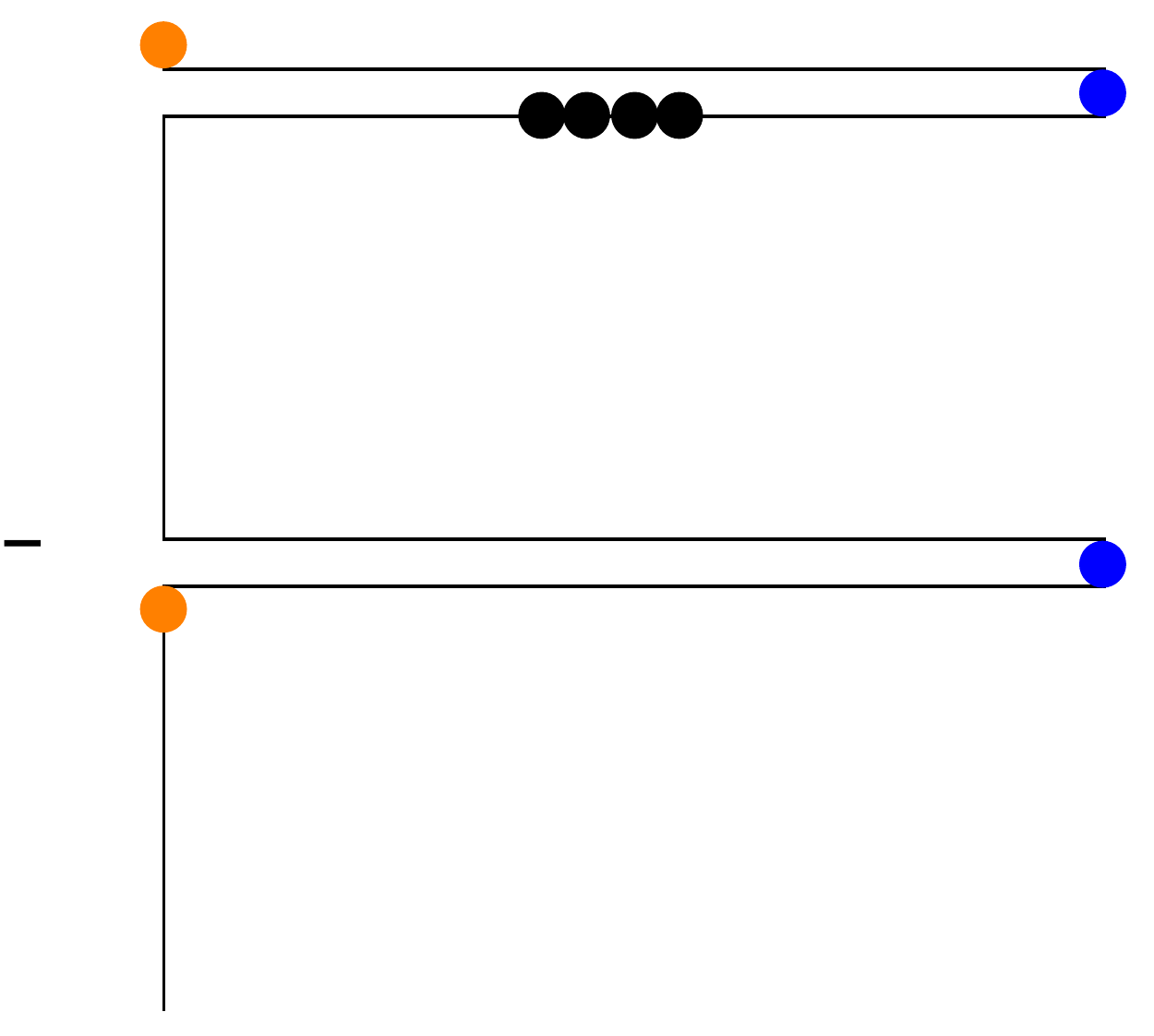}}\\
		\caption{The eight configurations of the commutator-squared OTOC. The four black dots represent the interaction vertex.}
		\label{8Configurations}
	\end{figure}
\end{center}

\section{The Lyapunov exponent: the vector case}\label{sec:LyapunovVector}
The Lyapunov exponent $\lambda_{L}$ comes from the late-time rate of growth of the $N^{0}$ term in $C(t)$. As explained in \cite{Stanford:2015owe}, to find $\lambda_{L}$ at leading order in the couplings, we proceed order by order in the coupling constants, and keep only the diagram which grows fastest in time at each order. Those diagrams turn out to be ladder diagrams, with the 1-rung diagram growing as $\lambda^{2} t$, the 2-rung diagram growing as $(\lambda^{2} t)^{2}$, etc. Then $C(\omega)$ is the sum of ladder diagrams. In particular, cross-ladder diagrams or double-ladder diagrams will be ignored, as in \cite{Stanford:2015owe}. The cross-ladders can be expected to give rise to subleading time dependences at late time, and the double-ladders give rise to exponential growth with a prefactor which is higher order in $1/N$ (See appendix \ref{app:DoubleLadders}). We will also ignore the integral of the interaction vertex along the vertical segments of the S-K contour, again like in \cite{Stanford:2015owe}, because such contributions only modify the thermal state without affecting the late-time growth.

Defining $f(\omega, p)$ by:
\begin{align}
	C(\omega) &= \frac1N \intmo{p} f(\omega, p)
\end{align}
Then $f(\omega, p)$ is given by the following sum of ladders:
\begin{equation}
	\begin{tikzpicture}[baseline=(current bounding box.center)]
		\begin{feynman}
			\node[blob] (i) at (0,0);
			\pgfmathsetmacro\vert{2*sin(pi / 6 r)};
			\pgfmathsetmacro\horz{2*cos(pi / 6 r)};
			\vertex [above right=\vert cm and \horz cm of i] (top);
			\vertex [below right=\vert cm and \horz cm of i] (bottom);
			
			\diagram* [] {
				(i) -- (top),
				(i) -- (bottom),
			};
		\end{feynman}
	\end{tikzpicture}
	=~
	\begin{tikzpicture}[baseline=(current bounding box.center)]
		\begin{feynman}
			\vertex(i);
			\pgfmathsetmacro\vert{sin(pi / 6 r)};
			\pgfmathsetmacro\horz{cos(pi / 6 r)};
			\vertex [above right=\vert cm and \horz cm of i] (looptop);
			\vertex [below right=\vert cm and \horz cm of i] (loopbottom);
			\vertex [above right=\vert cm and \horz cm of looptop] (top);
			\vertex [below right=\vert cm and \horz cm of loopbottom] (bottom);
			
			\diagram* [] {
				(i) -- (looptop) -- (top),
				(i) -- (loopbottom) -- (bottom),
			};
		\end{feynman}
	\end{tikzpicture}
	+~
	\begin{tikzpicture}[baseline=(current bounding box.center)]
		\begin{feynman}
			\vertex(i);
			\pgfmathsetmacro\vert{sin(pi / 6 r)};
			\pgfmathsetmacro\horz{cos(pi / 6 r)};
			\vertex [above right=\vert cm and \horz cm of i] (looptop);
			\vertex [below right=\vert cm and \horz cm of i] (loopbottom);
			\vertex [above right=\vert cm and \horz cm of looptop] (top);
			\vertex [below right=\vert cm and \horz cm of loopbottom] (bottom);
			
			\diagram* [] {
				(i) -- (looptop) -- (top),
				(i) -- (loopbottom) -- (bottom),
				(looptop) -- [out = -45, in = 45] (loopbottom),
				(looptop) -- [out = -135, in = 135] (loopbottom),
			};
		\end{feynman}
	\end{tikzpicture}
	+~
	\begin{tikzpicture}[baseline=(current bounding box.center)]
		\begin{feynman}
			\vertex(i);
			\pgfmathsetmacro\vert{sin(pi / 6 r)};
			\pgfmathsetmacro\horz{cos(pi / 6 r)};
			\vertex [above right=\vert cm and \horz cm of i] (looptop);
			\vertex [below right=\vert cm and \horz cm of i] (loopbottom);
			\vertex [above right=\vert cm and \horz cm of looptop] (top);
			\vertex [below right=\vert cm and \horz cm of loopbottom] (bottom);
			
			\diagram* [] {
				(i) -- (looptop) -- (top),
				(i) -- (loopbottom) -- (bottom),
				(looptop) -- [out = -45, in = 135] (loopbottom),
				(looptop) -- [out = -135, in = 45] (loopbottom),
			};
		\end{feynman}
	\end{tikzpicture}
	+ \cdots +~
	\begin{tikzpicture}[baseline=(current bounding box.center)]
		\begin{feynman}
			\vertex(i);
			\pgfmathsetmacro\vert{2/3*sin(pi / 6 r)};
			\pgfmathsetmacro\horz{2/3*cos(pi / 6 r)};
			\vertex [above right=\vert cm and \horz cm of i] (loop1top);
			\vertex [below right=\vert cm and \horz cm of i] (loop1bottom);
			\vertex [above right=\vert cm and \horz cm of loop1top] (loop2top);
			\vertex [below right=\vert cm and \horz cm of loop1bottom] (loop2bottom);
			\vertex [above right=\vert cm and \horz cm of loop2top] (top);
			\vertex [below right=\vert cm and \horz cm of loop2bottom] (bottom);
			
			\diagram* [] {
				(i) -- (loop1top) -- (loop2top) -- (top),
				(i) -- (loop1bottom) -- (loop2bottom) -- (bottom),
				(loop1top) -- [out = -45, in = 45] (loop1bottom),
				(loop1top) -- [out = -135, in = 135] (loop1bottom),
				(loop2top) -- [out = -45, in = 45] (loop2bottom),
				(loop2top) -- [out = -135, in = 135] (loop2bottom),
			};
		\end{feynman}
	\end{tikzpicture}
	+~
	\begin{tikzpicture}[baseline=(current bounding box.center)]
		\begin{feynman}
			\vertex(i);
			\pgfmathsetmacro\vert{2/3*sin(pi / 6 r)};
			\pgfmathsetmacro\horz{2/3*cos(pi / 6 r)};
			\vertex [above right=\vert cm and \horz cm of i] (loop1top);
			\vertex [below right=\vert cm and \horz cm of i] (loop1bottom);
			\vertex [above right=\vert cm and \horz cm of loop1top] (loop2top);
			\vertex [below right=\vert cm and \horz cm of loop1bottom] (loop2bottom);
			\vertex [above right=\vert cm and \horz cm of loop2top] (top);
			\vertex [below right=\vert cm and \horz cm of loop2bottom] (bottom);
			
			\diagram* [] {
				(i) -- (loop1top) -- (loop2top) -- (top),
				(i) -- (loop1bottom) -- (loop2bottom) -- (bottom),
				(loop1top) -- [out = -45, in = 45] (loop1bottom),
				(loop1top) -- [out = -135, in = 135] (loop1bottom),
				(loop2top) -- [out = -45, in = 135] (loop2bottom),
				(loop2top) -- [out = -135, in = 45] (loop2bottom),
			};
		\end{feynman}
	\end{tikzpicture}
	+ \cdots \nonumber \\
\end{equation}
where frequency $\omega$ flows into each diagram at the left corner. Also, as drawn above, the 1-rung diagram splits into the sum over several rung shapes (2 of which are drawn explicitly above), and similarly for the higher-rung diagrams.

By noting that an infinite ladder remains the same if we add one more rung to it, we arrive at the relation:
\begin{equation}
	\begin{tikzpicture}[baseline=(current bounding box.center)]
		\begin{feynman}
			\node[blob] (i) at (0,0);
			\pgfmathsetmacro\vert{2*sin(pi / 6 r)};
			\pgfmathsetmacro\horz{2*cos(pi / 6 r)};
			\vertex [above right=\vert cm and \horz cm of i] (top);
			\vertex [below right=\vert cm and \horz cm of i] (bottom);
			
			\diagram* [] {
				(i) -- (top),
				(i) -- (bottom),
			};
		\end{feynman}
	\end{tikzpicture}
	=~
	\begin{tikzpicture}[baseline=(current bounding box.center)]
		\begin{feynman}
			\vertex(i);
			\pgfmathsetmacro\vert{sin(pi / 6 r)};
			\pgfmathsetmacro\horz{cos(pi / 6 r)};
			\vertex [above right=\vert cm and \horz cm of i] (looptop);
			\vertex [below right=\vert cm and \horz cm of i] (loopbottom);
			\vertex [above right=\vert cm and \horz cm of looptop] (top);
			\vertex [below right=\vert cm and \horz cm of loopbottom] (bottom);
			
			\diagram* [] {
				(i) -- (looptop) -- (top),
				(i) -- (loopbottom) -- (bottom),
			};
		\end{feynman}
	\end{tikzpicture}
	+~
	\begin{tikzpicture}[baseline=(current bounding box.center)]
		\begin{feynman}
			\node[blob] (i) at (0,0);
			\pgfmathsetmacro\vert{sin(pi / 6 r)};
			\pgfmathsetmacro\horz{cos(pi / 6 r)};
			\vertex [above right=\vert cm and \horz cm of i] (looptop);
			\vertex [below right=\vert cm and \horz cm of i] (loopbottom);
			\vertex [above right=\vert cm and \horz cm of looptop] (top);
			\vertex [below right=\vert cm and \horz cm of loopbottom] (bottom);
			
			\diagram* [] {
				(i) -- (looptop) -- (top),
				(i) -- (loopbottom) -- (bottom),
				(looptop) -- [out = -45, in = 45] (loopbottom),
				(looptop) -- [out = -135, in = 135] (loopbottom),
			};
		\end{feynman}
	\end{tikzpicture}
	+~
	\begin{tikzpicture}[baseline=(current bounding box.center)]
		\begin{feynman}
			\node[blob] (i) at (0,0);
			\pgfmathsetmacro\vert{sin(pi / 6 r)};
			\pgfmathsetmacro\horz{cos(pi / 6 r)};
			\vertex [above right=\vert cm and \horz cm of i] (looptop);
			\vertex [below right=\vert cm and \horz cm of i] (loopbottom);
			\vertex [above right=\vert cm and \horz cm of looptop] (top);
			\vertex [below right=\vert cm and \horz cm of loopbottom] (bottom);
			
			\diagram* [] {
				(i) -- (looptop) -- (top),
				(i) -- (loopbottom) -- (bottom),
				(looptop) -- [out = -45, in = 135] (loopbottom),
				(looptop) -- [out = -135, in = 45] (loopbottom),
			};
		\end{feynman}
	\end{tikzpicture}
	+ \cdots \\
\end{equation}
which amounts to an integral equation in $f(\omega, p)$:
\begin{align}
	f(\omega,p) &= -G_{R}(p)G_{R}(\omega - p) \bigg[ 1 + \intmo{k} R_{total}{(p,k)} f(\omega, k)\bigg]
\end{align}
where $R_{total}$ is the total rung function, or the sum over all rung shapes. We can drop the inhomogeneous term, because we do not expect it to affect the late-time behavior:
\begin{align}
	f(\omega, p) &= -G_{R}(p)G_{R}(\omega - p) \intmo{k} R_{total}{(p,k)} f(\omega, k)
\end{align}
As explained in \cite{Stanford:2015owe}, the product of the retarded correlators on the right-hand side above can be replaced by:
\begin{equation}
    G_{R}{(p)}G_{R}{(\omega-p)} \rightarrow -\frac{\pi i}{2 E_{\textbf{p}}^{2}} \frac{\delta(p^{0} - E_{\textbf{p}}) + \delta(p^{0} + E_{\textbf{p}})}{\omega + 2i\Gamma_{total,\textbf{p}}}
\end{equation}
where $\Gamma_{total,\textbf{p}}$ is the total width, which comes from the imaginary part of the self-energy correction to the propagator (the real part is a momentum-dependent mass shift and should not affect the Lyapunov exponent at leading order in the t'Hooft coupling). The integral equation then becomes:
\begin{equation}
	(-i\omega + 2\Gamma_{\bff{p}})f(\omega, p) = \frac{\pi}{E_{\bff{p}}} \delta(p_0^2 - E_{\bff{p}}^2) \intmo{k} R_{total}{(p,k)} f(\omega, k)
\end{equation}
From the delta functions in the pairs of retarded propagators, we see that $f(\omega,p)$ only has support on shell, so we write $f(\omega,p) = f(\omega,\textbf{p}) \delta{(p_{0}^{2} - E_{\textbf{p}}^{2})}$, and the integral equation above becomes:
\begin{equation}\label{IntEq}
    	-i\omega f(\omega, \bff{p}) = -2\Gamma_{\bff{p}} f(\omega, \bff{p}) + \int\frac{d^3k}{(2\pi)^3} m(\bff{k}, \bff{p}) f(\omega, \bff{k})
\end{equation}
with
\begin{equation}
    m(\textbf{k},\textbf{p}) = \frac{R_{total}{(-E_{\textbf{p}},\textbf{p};E_{\textbf{k}},\textbf{k})} + R_{total}{(E_\textbf{p},\textbf{p};E_{\textbf{k}},\textbf{k})} }{4E_{\textbf{k}}E_{\textbf{p}}}
\end{equation}

We now describe in details the different contibutions to the total rung function $R_{total}$. It turns out that there are 7 different inequivalent shapes of the rung:
\begin{equation}
    R_{total}{(p,k)} = \sum_{i=1}^{7} R_{i}{(p,k)}
\end{equation}
as listed in Table \ref{table:ladders}, which shows the vertex factor for each of the 7 classes of diagram where no components of $\Theta$ have been set to zero for completeness. For each of the 7 classes of diagrams in that table, we have chosen to draw only one representative diagram belonging to the class, and we elaborate more on the remaining diagrams in the class in Appendix \ref{app:DiagramClasses}. Even though we have kept $\Theta^{0i} \neq 0$ in Table \ref{table:ladders}, everywhere else in the main body of this paper, $\Theta^{\mu 0}$, will be set to $0$ so that there is no non-commutativity in the temporal direction. 

In Table \ref{table:ladders}, $\ell$ is momentum running down one of the two Wightman functions making up the rung. To obtain the $R_{i}$'s, we first write down the integral over $\ell$, $\int \frac{d^{4} \ell}{(2\pi)^4} \tilde{G}{(p/2 + \ell)} \tilde{G}{(p/2 - \ell)}$.  We then multiply the integrand by the vertex factor, the prefactor and the number of diagrams per class, as listed in Table \ref{table:ladders}. We note that, when we set $\Theta^{\mu 0} = 0$, the 6th rung shape in the table is equivalent to the first one, so the total number of inequivalent rung shapes in that case is 6. We also note that, when the vertex factor is trivial, instead of working with $R_{i}$ as a function of 2 arguments, we can work with the difference between those two arguments and define $R_{i}$ to be a function of a single argument (the overall sign of that difference of the two arguments does not matter, because $\tilde{G}$ is even).

We also describe in details the various contributions to the total width $\Gamma_{total,\textbf{p}}$. The width comes from melon diagrams, and there are 6 inequivalent ones:
\begin{equation}
    \Gamma_{total,\textbf{p}} = \sum_{i=1}^{6} \Gamma_{i,\textbf{p}} 
\end{equation}
We list in Table \ref{table:melons} below the 6 types of melon diagrams, including the vertex factor and prefactor (which includes the dependence on the couplings as well as any combinatoric factor) for each diagram.

Solving the integral equation (\ref{IntEq}) is quite nontrivial, even numerically. Therefore, we will content ourselves with solving it in two limiting cases: the commutative case (where $\Theta^{\mu\nu} = 0$), and the large $\Theta^{\mu\nu}$ case.

\subsection{The commutative case} \label{sec:commutativeCase}
Let's compute the total rung function in the commutative case. All the vertex factors in Table \ref{table:ladders} simplify to $1$ in this case. Also, the two couplings become the same, so we can set $\lambda_{1} = \lambda_{2} \equiv \lambda$. Summing up all the rows of Table 1 then yields:
\begin{equation}
    R_{total}(p) = \frac{24\lambda^{2}}{N} \int \frac{d^{4}l}{(2\pi)^4} \tilde{G}{\left(\frac{p}{2} + l \right)} \tilde{G}{\left( \frac{p}{2} - l \right)}
\end{equation}
The integral over $l$ can be evaluated using the same method as \cite{Stanford:2015owe}. After plugging in a Wightman function of the free theory,
\begin{equation}
    \tilde{G}{(k)} = \sum_{s = \pm 1} \frac{\pi \delta(k^{0} - sE_{\textbf{k}})}{2E_{\textbf{k}} \sinh{(\beta E_{\textbf{k}}/2)}},
\end{equation}
the total rung function becomes,
\begin{equation}\label{3point13}
    R_{total}{(p)} = \frac{3\lambda^{2}}{8\pi^2 N} \int d^{3}l~ \frac{\delta(|p^{0}|-E_{-}-E_{+}) + 2\delta(|p^{0}| + E_{+} - E_{-})}{E_{+}E_{-} \sinh{(\beta E_{+}/2)} \sinh{(\beta E_{-}/2)}}.
\end{equation}
where $E_{\pm} \equiv E_{\frac{\textbf{p}}{2} \pm l}$. The integral is then easiest to evaluate if we change integration variables from Cartesian components of $l$ to $E_{\pm}$. We note that $E_{\pm}$ are functions of the component of $l$ parallel to $\textbf{p}$, and the magnitude of the perpendicular part of $l$ to $\textbf{p}$ (but not of the azimuthal angle around $\textbf{p}$). Also, the domain of integration in the $(E_{+},E_{-})$ plane can be found by mapping from the $l$-space, and is found to be the region satisfying the inequality,
\begin{equation}
    E_{+}^{2} + E_{-}^{2} \geq 2m^{2} + \frac{|\textbf{p}|^2}{2} + \frac{1}{2|\textbf{p}|^2} (E_{+}^{2} - E_{-}^{2})^{2}.
\end{equation}
In addition, the support of the delta functions in (\ref{3point13}) will be straight lines in the $(E_{+},E_{-})$ plane. By working out the intersections between those lines and the region described by the inequality above, we can evaluate the integral in (\ref{3point13}), and find,
\begin{equation}
    R_{total}{(p)} = \frac{3\lambda^{2}}{\pi\beta N} \frac{1}{|\textbf{p}|\sinh{(\beta |p^{0}|/2)}} \bigg[ \theta(-p^{2}-4m^{2}) \log\frac{\sinh x_{+}}{\sinh x_{-}} + \theta(p^{2}) \log \frac{1-e^{-2x_{+}}}{1-e^{2x_{-}}} \bigg],
\end{equation}
with
\begin{equation}
    x_{\pm} = \frac{\beta}{4} \left( |p^{0}| \pm |\textbf{p}| \sqrt{1 + \frac{4m^2}{\textbf{p}^{2} - (p^{0})^{2}}} \right).
\end{equation}

Next, we need the total width. Again, all the vertex factors in Table \ref{table:melons} simplify to $1$. Setting $\lambda_{1} = \lambda_{2}$, and adding up the rows in Table \ref{table:melons}, we find the 2-loop self-energy (in Euclidean signature) to be:
\begin{equation}
    \Pi{(i\omega_{n})} = \frac{-8\lambda^{2}}{N} \int_{0}^{\infty} d\tau e^{i\omega_{n}\tau} G(\tau)^{3}
\end{equation}
In terms of the spectral function $\rho$, the Euclidean correlator is,
\begin{equation}
    G(\tau) = \int \frac{dk^{0}}{2\pi} \frac{\rho(k^{0}) e^{-k^{0}\tau}}{1 - e^{-\beta k^{0}}},
\end{equation}
and the 2-loop self-energy takes the form,
\begin{equation}
    \Pi{(i\omega_{n})} = \frac{8\lambda^{2}}{N} \int \Pi_{j=1}^{3} \bigg[ \frac{dk^{0}_{j}}{2\pi} \frac{\rho(k^{0}_{j})}{1 - e^{-\beta k^{0}_{j}}} \bigg] \frac{1-e^{-(k^{0}_{1}+k^{0}_{2}+k^{0}_{3})\beta}}{i\omega_{n} - (k^{0}_{1} + k^{0}_{2} + k^{0}_{3})}.
\end{equation}
Continuing $i\omega_{n} \rightarrow p^{0} + i\epsilon$, and taking the imaginary part:
\begin{eqnarray}
    -\mathrm{Im}{[\Pi(p^{0}+i\epsilon, \textbf{p})]} &=& \frac{8\lambda^{2}}{N} \sinh{\frac{\beta p^{0}}{2}} \int \Pi_{j=1}^{3} \bigg[ \frac{d^{4}k_{j}}{(2\pi)^4} \frac{\rho(k_{j})}{2\sinh{(\beta k^{0}_{j}/2)}} \bigg] (2\pi)^{4} \delta^{4}{(p-k_{1}-k_{2}-k_{3})} \nonumber \\
    &=& \frac{8\lambda^{2}}{N} \sinh{\frac{\beta p^{0}}{2}} \int \frac{d^{4}k_1}{(2\pi)^4} \int \frac{d^{4}k_2}{(2\pi)^4} \tilde{G}{(k_1)} \tilde{G}{(k_2)} \tilde{G}{(p-k_{1}-k_{2})}.
\end{eqnarray}
In the second equality, we used a relation between the Wightman and spectral density functions given by,
\begin{equation}
    \tilde{G}{(k)} = \frac{\rho(k)}{2\sinh{(\beta k^{0}/2)}}.
\end{equation}
The width is related to the imginary part of the self-energy correction by $\Gamma_{total, \textbf{p}} = -\mathrm{Im}{(E_{\textbf{p}}+i\epsilon, \textbf{p})}/2E_{\textbf{p}}$. We then find that the total width can be written in terms of the total rung function by,
\begin{eqnarray}
    \Gamma_{total,\textbf{p}} &=& \frac{\sinh{(\beta E_{\textbf{p}}/2)}}{6E_{\textbf{p}}} \int \frac{d^{4}k}{(2\pi)^4} R_{total}{(p-k)} \tilde{G}{(k)} |_{p^{0} = E_{\textbf{p}}} \nonumber \\
    &=& \frac{1}{6} \int \frac{d^{3}k}{(2\pi)^3} \frac{\sinh{(\beta E_{\textbf{p}}/2)}}{\sinh{(\beta E_{\textbf{k}}/2)}} m(\textbf{k}, \textbf{p}),
\end{eqnarray}
where we used the explicit form of $\tilde{G}$ in the second equality.

The integral equation can then be brought to the form,
\begin{equation}
    -i\omega f(\omega,\textbf{p}) = \int \frac{d^{3}k}{(2\pi)^3} m(\textbf{k},\textbf{p}) \bigg( f(\omega,\textbf{k}) - \frac{\sinh{(\beta E_{\textbf{p}}/2)}}{3 \sinh{(\beta E_{\textbf{k}}/2)}} f(\omega,\textbf{p}) \bigg).
\end{equation}
This equation has the form of a first order differential equation in real time, $\dv{f}{t} = M f$, where $M$ is the integral operator on the right hand side \cite{Stanford:2015owe}.  The largest positive eigenvalue of the operator $M$ characterizes the exponential growth of the OTOC in question. To compute these eigenvalues numerically, we first note that in the commutative case, we have spherical symmetry, so that $f$ depends on $\textbf{p}$ only through its norm $|\textbf{p}|$. Also, $m(\textbf{k},\textbf{p})$ depends on the two vectors in the arguments through their norms and $y \equiv |\textbf{k} - \textbf{p}|$. We can also change integration variables to $|\textbf{k}|$ and $y$ (note that the integrand is independent of the azimuthal angle around $\textbf{p}$). The integral equation then takes the form,
\begin{equation} \label{eq:IntEqSpherical}
    -i\omega f(\omega,|\textbf{p}|) = \int_{0}^{\infty} d|\textbf{k}|~ m_{1}{(|\textbf{k}|,|\textbf{p}|)} \bigg( f(\omega,|\textbf{k}|) - \frac{\sinh(\beta E_{\textbf{p}}/2)}{3 \sinh{(\beta E_{\textbf{k}}/2)}} f(\omega,|\textbf{p}|) \bigg),
\end{equation}
with
\begin{equation}\label{m1}
    m_{1}{(|\textbf{k}|,|\textbf{p}|)} \equiv 2\pi \frac{|\textbf{k}|}{|\textbf{p}|} \int_{||\textbf{k}|-|\textbf{p}||}^{|\textbf{k}|+|\textbf{p}|} ydy~ m(|\textbf{k}|,|\textbf{p}|,y).
\end{equation}
The right-hand side of (\ref{eq:IntEqSpherical}) can then be discretized. The function $m_{1}$ becomes a matrix, with its two arguments thought of as the two indices of the matrix. We can then diagonalize that matrix and look for the largest positive eigenvalue, which is the Lyapunov exponent $\lambda_{L}$. Also, for numerical purposes, it is convenient to perform one more change of variable so that the  semi-infinite integration range over $\abs{\textbf{k}}$ is compactified into the interval $[0,1]$.

In the end, we find the Lyapunov exponent to be:
\begin{equation} \label{eq:lyapunovCommVector}
    \lambda_{L}^{C} = 0.0123 \frac{\lambda^2}{\beta^{2} m N}
\end{equation}

\subsection{The large Moyal-scale case}
In the limit where the Moyal area is larger than any other scale, most of the vertex factors of ladder diagrams become highly oscillatory as a function of $l$. Because the integral of such a highly oscillatory function tends to zero, ladder diagrams with $l$-dependent vertex factors can be ignored. Furthermore, the fifth ladder in Table \ref{table:ladders} has an $l$-independent vertex factor, so it does not tend to zero.  However, the additional integration in the integral equation will send this ladder to $0$. In this way, we see that only the first and sixth ladder diagrams (i.e. the ones with a trivial vertex factors) contribute in this limit.  Thus, the total rung function is,
\begin{equation} \label{eq:NCVectorRung}
    R_{\text{total}}{(p)} = \frac{3}{N} \left(\frac{1}{2}\lambda_{1}^{2} + \lambda_{2}^{2} \right) \int \frac{d^{4}l}{(2\pi)^4} ~\tilde{G}{\left(\frac{p}{2} + l \right)} \tilde{G}{\left(\frac{p}{2} - l\right)}.
\end{equation}
The integral over $l$ can be evaluated in the same way as the commutative limit.

Similarly, most of the melon diagrams' vertex factors tend to zero due to the same highly oscillatory behavior.  Only the planar melon contributes in the large Moyal scale limit. The 2-loop self-energy is then,
\begin{equation}
    \Pi{(i\omega_{n})} = -\frac{1}{N}\left(\frac{1}{2}\lambda_{1}^{2} + \lambda_{2}^{2} \right) \intl{0}{\infty} d\tau ~e^{i\omega_{n}\tau} G(\tau)^{3}.
\end{equation}
The total width can still be expressed in terms of the total rung function by the same relation as in the commutative case,
\begin{eqnarray}
    \Gamma_{total,\textbf{p}} &=& \frac{1}{6} \int \frac{d^{3}k}{(2\pi)^3} ~\frac{\sinh{(\beta E_{\textbf{p}}/2)}}{\sinh{(\beta E_{\textbf{k}}/2)}} m(\textbf{k}, \textbf{p}),
\end{eqnarray}
with $m$ still given by
\begin{equation} \label{eq:rungM}
    m(\textbf{k},\textbf{p}) = \frac{R_{total}{(-E_{\textbf{p}},\textbf{p};E_{\textbf{k}},\textbf{k})} + R_{total}{(E_\textbf{p},\textbf{p};E_{\textbf{k}},\textbf{k})} }{4E_{\textbf{k}}E_{\textbf{p}}}.
\end{equation}
The integral equation takes the same form as in the commutative case,
\begin{equation} \label{eq:integralEq}
    -i\omega f(\omega,\textbf{p}) = \int \frac{d^{3}k}{(2\pi)^3} ~m(\textbf{k},\textbf{p}) \bigg( f(\omega,\textbf{k}) - \frac{\sinh{(\beta E_{\textbf{p}}/2)}}{3 \sinh{(\beta E_{\textbf{k}}/2)}} f(\omega,\textbf{p}) \bigg),
\end{equation}
but there are minor differences: First, the overall factor in $m(\textbf{k},\textbf{p})$, which encodes the dependence on the couplings, as well as combinatoric factors, differs between the commutative limit and the large non-commutativity limit. Secondly, in the large non-commutativity limit, we do not have spherical symmetry, and we cannot take $f(\omega, \textbf{p})$ to be a function of $|\textbf{p}|$ only. We can, however, choose $\Theta^{12} = -\Theta^{21} = \Theta$ to be the only non-vanishing component of $\Theta^{\mu\nu}$. We then have axisymmetry around an axis perpendicular to the $1$-$2$ plane where we can work in cylindrical coordinates $(\rho, \phi, z)$.

Now, $f$ depends on $\textbf{p}$ through the components $p^{\rho}$ and $p^{z}$, but not $p^{\phi}$. The idea is then to think about the two arguments ($p^{z}$, $p^{\rho}$) of $f$ as a ``super-index'' which plays the role of the norm $|\textbf{p}|$ in the spherically symmetric case considered in the previous subsection. The right-hand side of the integral equation (\ref{eq:integralEq}) then takes the form of matrix multiplication between some matrix $m_{1}$ and $f$. This is just like in the spherically symmetric case, but with each index of the matrix $m_{1}$ now taken to be the super-index $(p^{z},p^{\rho})$. Furthermore, $m_{1}$ is itself an integral over the azimuthal component $p^{\phi}$ (just like the $m_{1}$ in (\ref{eq:integralEq}) is an integral over $y$).

We can then do a change of variables to compactify the integration domain, discretize the integral equation and find the Lyapunov spectrum. In the end, we find for the Lyapunov exponent,
\begin{equation}
    \lambda_{L}{(\Theta \rightarrow \infty)} = 0.00244~ \frac{\lambda_{1}^{2} + 2\lambda_{2}^{2}}{3\beta^{2} m N}.
\end{equation}

There is some question about what is being kept fixed when we compare Lyapunov exponents between theories with different $\Theta$.  In the analysis above, we have chosen to keep the couplings $\lambda_1$ and $\lambda_2$ fixed.  Of course there are other choices we could make, such as keeping the decay rate of the two-point function fixed.  However, we do not see a preferred choice of what is kept fixed, and we do not expect this choice to have a significant impact on the Lyapunov exponent's numerical value.

\subsection{A closer look at the highly oscillatory rung functions}

In this subsection, we take a closer look at the the claim that the nonplanar rung functions drop out at large $\Theta$. We will check this claim for the second rung function $R_{2}$ listed in Table \ref{table:ladders}, with the other rung functions being similarly treated.

We have,
\begin{equation}
    R_{2}{(p)} = \frac{1}{64 \pi^2} \int d^{3}l~ e^{i \textbf{l} \times \textbf{p}} ~ \frac{\delta(|p^{0}| - E_{+} - E_{-}) + 2\delta(|p^{0}| + E_{+} - E_{-})}{E_{+}E_{-}\sinh{\left(\frac{\beta E_{+}}{2} \right)} \sinh{\left(\frac{\beta E_{-}}{2} \right)}},
\end{equation}
with $E_{\pm} = E_{\frac{\textbf{p}}{2} \pm \textbf{l}}$. We distinguish two cases for the vector $p$: the generic case, where the vertex factor $e^{i \textbf{l} \times \textbf{p}}$ is nontrivial, and the nongeneric case where $\Theta \cdot \textbf{p}$ vanishes, and the vertex factor goes to unity. For a nongeneric value of $p$, $R_{2}(p)$ is independent of $\Theta$. However, this case can be neglected since the integral equation contains an integral over $p$, and the nongeneric values of $p$ form a set of measure zero, and hence do not contribute to the integral. An exception to this case would be if $R_{2}(p)$ were to become singular at those nongeneric values of $p$.  If $R_2$ became singular, then those particular values would need to be treated carefully, however $R_{2}(p)$ simply reduces to the planar rung function at the nongeneric $p$ in question (up to an overall factor).  Hence, there are no singular cases to treat individually.

We can now orient the vector $\textbf{p}$ along the 3-axis, and transform the measure of integration from $(l^{1},l^{2},l^{3})$ to $(E_{+},E_{-},\phi)$ (where $\phi$ is the angle ranging from $0$ to $2\pi$ running around the vector $\textbf{p}$). The rung function then becomes,
\begin{equation}
    R_{2}{(p)} = \frac{1}{64 \pi^2 |\textbf{p}|} \iint dE_{+} dE_{-} \frac{\delta(|p^{0}| - E_{+} - E_{-}) + 2\delta(|p^{0}| + E_{+} - E_{-})}{\sinh{\left(\frac{\beta E_{+}}{2} \right)} \sinh{\left(\frac{\beta E_{-}}{2} \right)}} \int_{0}^{2\pi} d\phi e^{i \textbf{l} \times \textbf{p}},
\end{equation}
where the integration domain in the $(E_{+},E_{-})$ plane is the same as for the planar rung function; i.e. the region satisfying the inequality,
\begin{equation}
    E_{+}^{2} + E_{-}^{2} \geq 2m^{2} + \frac{|\textbf{p}|^2}{2} + \frac{1}{2|\textbf{p}|^2} (E_{+}^{2} - E_{-}^{2})^{2}.
\end{equation}

The $\phi$-integral can be evaluated in terms of Bessel functions. We can use the delta functions to perform the $E_{+}$-integral. Breaking the rung function into the sum of two parts,
\begin{equation}
    R_{2}{(p)} = I_1 + I_2,
\end{equation}
where,
\begin{align}
    I_1 &= \frac{1}{32\pi|\textbf{p}|} \theta{(-p^{2}-4m^{2})} \int_{x_{-}}^{x_{+}} dE_{-} \frac{J_{0}{\bigg[ l_{\perp}{(|p^{0}|-E_{-},E_{-})} |\Theta \cdot \textbf{p}| \bigg]}}{\sinh{\left( \frac{\beta E_{-}}{2} \right)} \sinh{\left(\frac{\beta}{2} (|p^{0}|-E_{-}) \right)}} \\
    I_2 &= \frac{1}{16\pi|\textbf{p}|} \theta{(p^2)}
    \int_{x_{+}}^{\infty} dE_{-} \frac{J_{0}{\bigg[ l_{\perp}{(E_{-}-|p^{0}|,E_{-})} |\Theta \cdot \textbf{p}| \bigg]}}{\sinh{\left( \frac{\beta E_{-}}{2} \right)} \sinh{\left(\frac{\beta}{2} (E_{-}-|p^{0}|) \right)}},
\end{align}
with,
\begin{equation}
    x_{\pm} = \frac{1}{2} \left( |p^{0}| \pm |\textbf{p}| \sqrt{1 + \frac{4m^2}{|\textbf{p}|^{2} - (p^{0}){}^{2}}} \right),
\end{equation}
and $l_{\perp}(E_{+},E_{-})$ being the magnitude of the part of $\textbf{l}$ perpendicular to $\textbf{k}$, written as a function of $E_{+}$ and $E_{-}$. Explicitly,  
\begin{align}
	l_{\perp}{(E_{+},E_{-})} \equiv \sqrt{\frac{E_{+}^{2}+E_{-}^{2}}{2} - m^{2} - \frac{|\textbf{p}|^2}{4} - \frac{(E_{+}^{2}-E_{-}^{2})^2}{4|\textbf{p}|^{2}}}.
\end{align}
At large $\Theta$, we can approximate the Bessel function by its asymptotic form, and obtain,
\begin{align}
    I_1 &=  \theta{(-p^{2}-4m^{2})} \sqrt{\frac{2}{\pi}}~\Re \bigg[  \frac{e^{-i\pi/4}}{32\pi |\textbf{p}|} \times \nonumber\\
    &\hspace{.1\linewidth} \times \int_{x_{-}}^{x_{+}} dE_{-}  \frac{1}{\sqrt{|\Theta \cdot \textbf{p}| l_{\perp}{(E_{+},E_{-})}}} \frac{ e^{i|\Theta \cdot \textbf{p}| l_{\perp}{(E_{+},E_{-})}}}{\sinh{\left( \frac{\beta E_{-}}{2} \right)} \sinh{\left(\frac{\beta E_{+}}{2} \right)}} \bigg]_{E_{+} = |p^{0}| - E_{-}} \\
    I_2 &=   \theta{(p^{2})} \sqrt{\frac{2}{\pi}}~ \Re \bigg[  \frac{e^{-i\pi/4}}{16\pi |\textbf{p}|} \times \nonumber \\
    &\hspace{.05\linewidth} \times \int_{x_{+}}^{\infty} dE_{-}  \frac{1}{\sqrt{|\Theta \cdot \textbf{p}| l_{\perp}{(E_{+},E_{-})}}} \frac{ e^{i|\Theta \cdot \textbf{p}| l_{\perp}{(E_{+},E_{-})}}}{\sinh{\left( \frac{\beta E_{-}}{2} \right)} \sinh{\left(\frac{\beta E_{+}}{2} \right)}} \bigg]_{E_{+} = E_{-} - |p^{0}|}
\end{align}
It is quite easy to see that the expressions above vanish when $\Theta \rightarrow \infty$. If needed, we can use the stationary phase approximation to do the integrals above at large $\Theta$. We would then need to look for the extrema of the function  $l_{\perp}{(|p^{0}|-E_{-},E_{-})}$ in the phase of the exponential. This function has one critical point at $E_{-} = \frac{|p^{0}|}{2}$, which lies between the two limits of integration of $I_1$, but outside of the integration range of $I_2$.

\section{The Lyapunov exponent: the matrix case}\label{sec:LyapunovMatrix}
We now move on to discuss the Lyapunov exponent for the non-commutative matrix model. For the purposes of $N$-counting, it is convenient to draw diagrams using double-line notation \cite{Minwalla:1999px}. The two interaction vertices, $\Phi^a_b \star \Phi^b_c \star \Phi^c_d \star \Phi^d_a$ and $\Phi^a_b \star \Phi^b_c \star \Phi^d_a \star \Phi^c_d$, correspond to two different drawings in the double-line notation, as shown in Fig. \ref{fig:DoubleLineVertices} below.

\begin{center}
\begin{figure}[h!]
\centering
\includegraphics[width = 4in]{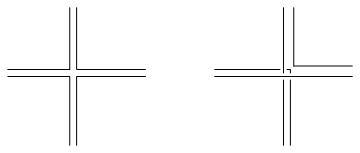}
\caption{Left: the quartic vertex $\Phi^a_b \star \Phi^b_c \star \Phi^c_d \star \Phi^d_a$ in the double-line notation. Right: the ``mixed-up'' matrix multiplication $\Phi^a_b \star \Phi^b_c \star \Phi^d_a \star \Phi^c_d$ in the double-line notation.}
\label{fig:DoubleLineVertices}
\end{figure}
\end{center}
\begin{figure}[t!]
\begin{equation*}
\begin{tikzpicture}[baseline=(current bounding box.center)]
	\begin{feynman}
		\vertex (ul);
		\vertex [below=.1cm of ul] (ul2);
		\vertex [below=2cm of ul] (ll);
		\vertex [above=.1cm of ll] (ll2);
		\vertex [right=3cm of ul] (ur);
		\vertex [below=.1cm of ur] (ur2);
		\vertex [right=3cm of ll] (lr);
		\vertex [above=.1cm of lr] (lr2);
		
		\vertex [right=1.5cm of ul2] (mmu);
		\vertex [left=.1cm of mmu] (mlu);
		\vertex [right=.1cm of mmu] (mru);
		
		\vertex [right=1.5 cm of ll2] (mml);
		\vertex [left=.1cm of mml] (mll);
		\vertex [right=.1cm of mml] (mrl);
		
		\diagram*{
			(ul) -- (ur),
			(ur) -- [out=0, in=0, max distance=1cm] (lr),
			(lr) -- (ll),
			(ll) -- [out=180, in=180, max distance=1cm] (ul),
			
			(ul2) -- (mlu),
			(mlu) -- [out=-135, in=135, max distance=1cm] (mll),
			(mll) -- (ll2),
			(ll2) -- [out=180, in=180, max distance=.6cm] (ul2),
			
			(mmu) -- [out=-45, in=45, max distance=1cm] (mml),
			(mmu) -- [out=-135, in=135, max distance=1cm] (mml),
			
			(ur2) -- [out=0, in=0, max distance=.6cm] (lr2),
			(lr2) -- (mrl),
			(mrl) -- [out=45, in=-45, max distance=1cm] (mru),
			(mru) -- (ur2)
			
		};
	\end{feynman}
\end{tikzpicture}
{}~~~
\begin{tikzpicture}[baseline=(current bounding box.center)]
	\begin{feynman}
		\vertex (ul);
		\vertex [below=.1cm of ul] (ul2);
		\vertex [below=2cm of ul] (ll);
		\vertex [above=.1cm of ll] (ll2);
		\vertex [right=3cm of ul] (ur);
		\vertex [below=.1cm of ur] (ur2);
		\vertex [right=3cm of ll] (lr);
		\vertex [above=.1cm of lr] (lr2);
		
		\vertex [right=1.45cm of ul] (mulu);
		\vertex [right=.1cm of mulu] (muru);
		\vertex [below=.1cm of mulu] (mllu);
		\vertex [right=.1cm of mllu] (mlru);
		
		\vertex [right=1.45cm of ll] (mlll);
		\vertex [right=.1cm of mlll] (mlrl);
		\vertex [above=.1cm of mlll] (mull);
		\vertex [right=.1cm of mull] (murl);
		
		\diagram*{
			(ul2) -- (mllu),
			(mllu) -- (mull),
			(mull) -- (ll2),
			(ll2) -- [out=180,in=180, max distance =.6cm] (ul2),
			
			(mlll) -- (ll),
			(ll) -- [out=180,in=180, max distance=1cm] (ul),
			(ul) -- (mulu),
			(mulu) -- [out=80,in=-80, min distance=2cm] (mlll),
			
			(muru) -- [out=80,in=-80, min distance=2cm] (mlrl),
			(muru) -- (ur),
			(ur) -- [out=0,in=0, max distance=1cm] (lr),
			(lr) -- (mlrl),
			
			(mlru) -- (ur2),
			(ur2) -- [out=0,in=0, max distance = .6cm] (lr2),
			(lr2) -- (murl),
			(murl) -- (mlru)
		};
	\end{feynman}
\end{tikzpicture}
{}~~~
\begin{tikzpicture}[baseline=(current bounding box.center)]
	\begin{feynman}
		\vertex (ul);
		\vertex [below=.1cm of ul] (ul2);
		\vertex [below=2cm of ul] (ll);
		\vertex [above=.1cm of ll] (ll2);
		\vertex [right=3cm of ul] (ur);
		\vertex [below=.1cm of ur] (ur2);
		\vertex [right=3cm of ll] (lr);
		\vertex [above=.1cm of lr] (lr2);
		
		\vertex [right=1.45cm of ul] (mulu);
		\vertex [right=.1cm of mulu] (muru);
		
		\vertex [right=1.45cm of ll] (mlll);
		\vertex [right=.1cm of mlll] (mlrl);
		
		\vertex [above right = .3cm and 1.5cm of ul] (mu);
		\vertex [below right = .3cm and 1.5cm of ll] (ml);
		
		\diagram*{
			(ul2) -- (ur2),
			(ur2) -- [out=0,in=0, max distance = .6cm] (lr2),
			(lr2) -- (ll2),
			(ll2) -- [out=180,in=180, max distance =.6cm] (ul2),
			
			(ul) -- (mulu),
			(mulu) -- [out=100,in=-100, min distance = 2cm] (mlll),
			(mlll) -- (ll),
			(ll) -- [out=180,in=180, max distance = 1cm] (ul),
			
			(muru) -- [out=80, in=-80, min distance =2cm] (mlrl),
			(mlrl) -- (lr),
			(lr) -- [out=0,in=0,max distance = 1cm] (ur),
			(ur) -- (muru),
			
			(mu) -- [out=80,in=-80, min distance=1.5cm] (ml),
			(mu) -- [out=100,in=-100, min distance=1.5cm] (ml),
		};
	\end{feynman}
\end{tikzpicture}
\end{equation*}
\centering
\includegraphics[scale=.5]{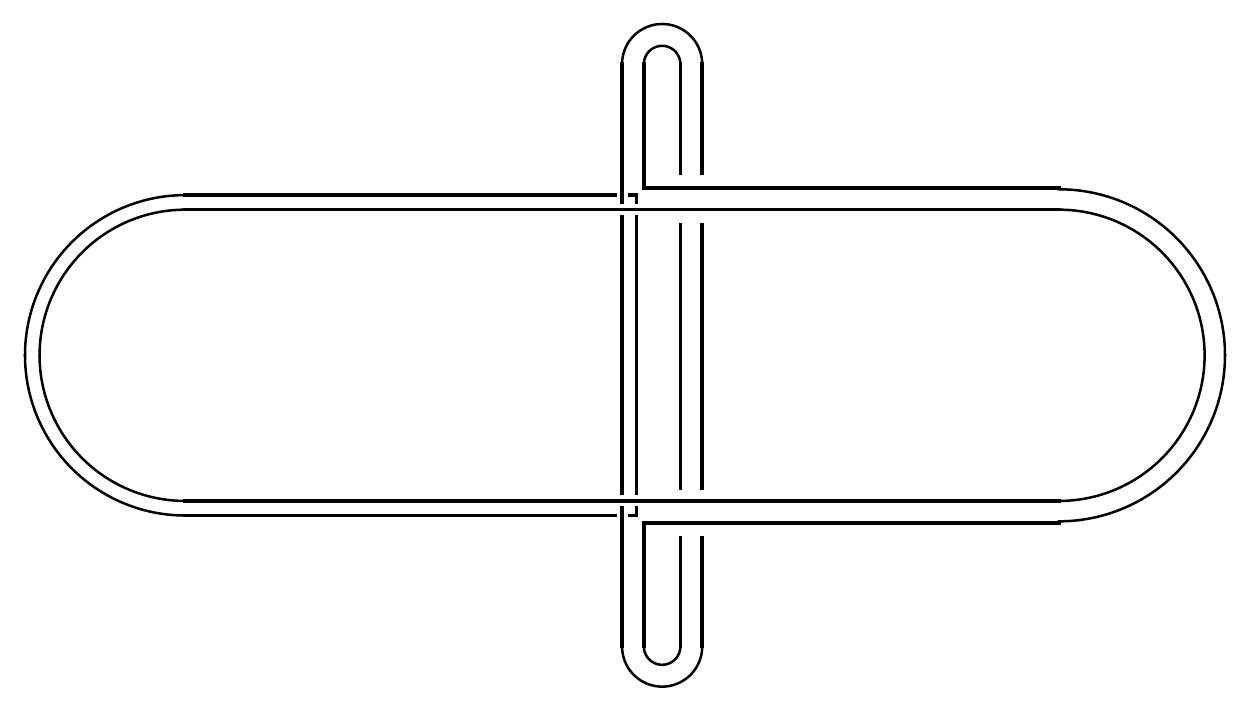}
\caption{Top: The three diagrams on top arise from the standard vertex shown in Fig. \ref{fig:DoubleLineVertices}.  Each of these diagrams has a counterpart, but with the standard vertex switched with the mixed up vertex.  Bottom: A representative diagram from the mixed up vertex diagrams is shown.  This is the 1-hump mixed up vertex diagram.  There are four leading order in $N$ diagrams like this one, and the same goes for the standard and 4-hump diagrams.}
\label{fig:doubleLineDiagrams}
\end{figure}

As in the vector case, we'll need to work out the combinatoric prefactors and vertex factors of each ladder diagram and melon diagram drawn in Tables \ref{table:ladders} and \ref{table:melons}. The vertex factors will be the same as for the vector case, and only the combinatoric prefactors (which include the dependence on the couplings $\lambda_1$ and $\lambda_2$) will be different. As in the vector case, we will only be interested in the large $\Theta$ limit, so it will be enough to find the prefactors for those diagrams which survive in this limit. Among the ladder diagrams, the surviving ones are those in the first class and the next-to-last one listed in Table \ref{table:ladders} (i.e. the planar diagram, the ``4-hump'' diagram, and the ``2-hump'' diagram). Among the melon diagrams, only the planar one (the first in Table \ref{table:melons}) survives the large $\Theta$ limit.

Let's now count the number of Wick contractions which are leading order in $1/N$ for each of the 4 surviving diagrams mentioned above. For the planar ladder diagram, we have 3 possibilities: 
\begin{enumerate}
	\item Both interaction vertices are $\Phi^a_b \star \Phi^b_c \star \Phi^c_d \star \Phi^d_a$ (so they both look like the left vertex in Fig. \ref{fig:DoubleLineVertices}).
	\item Both are $\Phi^a_b \star \Phi^b_c \star \Phi^d_a \star \Phi^c_d$ (so they both look like the right vertex in figure \ref{fig:DoubleLineVertices}).
	\item One of them is $\Phi^a_b \star \Phi^b_c \star \Phi^c_d \star \Phi^d_a$ and the other one is $\Phi^a_b \star \Phi^b_c \star \Phi^d_a \star \Phi^c_d$.
\end{enumerate}

The first possibility gives rise to 16 leading-order Wick contractions (where $16 =4 \times 4$ comes from the freedom to rotate each of the double-line vertices in the plane of the diagram). The second possibility gives rise to 4 leading-order contractions (as mentioned in Fig. \ref{fig:doubleLineDiagrams}).  These are the four double-line diagrams which are reflection symmetric between the two rails of the ladders. The third possibility gives rise to no leading-order contractions.

The next two types of diagrams, the ``4-hump'' and ``2-hump'' diagrams, can be analyzed in a manner completely analogous to the planar diagram, and in fact the combinatorics for combinations of interaction terms in the list above are exactly the same.  Hence, the combinatoric prefactor associated with the ladder diagrams' total rung function is given by,
\begin{align}
	3\frac{16 \times 2}{4^2 \times 2}~\lambda_1^2 + 3\frac{4\times 2}{4^2 \times 2}~\lambda_2^2 = 3\lambda_1^2 + \frac34 \lambda_2^2.
\end{align}
The first term in the sum comes from the diagrams where both interaction vertices are $\Phi^a_b \star \Phi^b_c \star \Phi^d_a \star \Phi^c_d$, while the second term in the sum comes from the diagrams where both interaction vertices are $\Phi^a_b \star \Phi^b_c \star \Phi^d_a \star \Phi^c_d$.  Focusing on the first term: the $3$ is due to there being three types of diagrams which contribute at leading order; the $16$ comes from the $16$ leading order Wick contractions; the $2$ in the numerator is from the freedom to interchange the two vertices; the $4^2$ comes from the interaction coupling, $\lambda_1/4$, at second order; and finally the $2$ in the denominator comes from the $1/2$ in the Dyson series expansion.  Now turning our attention to the second term: the $3$ is due to there being three types of diagrams which contribute at leading order; the $4$ comes from the $4$ leading order Wick contractions; the $2$ in the numerator is from the freedom to interchange the two vertices; the $4^2$ comes from the interaction coupling, $\lambda_2/4$, at second order; and finally the $2$ in the denominator comes from the $1/2$ in the Dyson series expansion.  This can be compared to the combinatoric factor of $12$ found in \cite{Stanford:2015owe} where the theory was commutative.

We are now in a position to write down the equivalent of (\ref{eq:NCVectorRung}) for the non-commutative matrix model at large $\Theta$.  It is given by
\begin{align} \label{eq:NCMatrixRung}
	R_{\text{matrix}}(p) &= 3\left(\lambda_1^2 + \frac14 \lambda_2^2\right) \intmo{l} \tilde{G}\left(\frac{p}{2} + l\right) \tilde{G}\left(\frac{p}{2} - l\right).
\end{align}

We can now compute the combinatoric factor of the melon diagram that contributes to the two loop width (imaginary part of the self-energy).  As with the ladder diagrams, there are three possible combinations of interaction vertices.  Using the same list as above, scenario $1$ provides $16$ leading order Wick contractions from the freedom to rotate each of the double-line vertices.  Scenario $2$ gives rise to $4$ leading order Wick contractions while scenario $3$ has no leading order Wick contractions in $\Theta$.  Hence the total combinatoric factor is
\begin{align}
	\frac{16\times 2}{4^2\times2} ~\lambda_1^2 + \frac{4\times2}{4^2\times2} ~\lambda_2^2 &= \lambda_1^2 + \frac14\lambda_2^2.
\end{align}
This yields a self-energy and width of (after following the same procedures as given in section \ref{sec:commutativeCase}),
\begin{align}
	\Pi{(i\omega_{n})} &= -\left(\lambda_1^2 + \frac14\lambda_2^2\right)  \int_{0}^{\infty} d\tau e^{i\omega_{n}\tau} G(\tau)^{3}\\
	\Gamma_{\bff{p}} &= \frac{1}{6} \int \frac{d^{3}k}{(2\pi)^3} ~\frac{\sinh{(\beta E_{\textbf{p}}/2)}}{\sinh{(\beta E_{\textbf{k}}/2)}} m(\textbf{k}, \textbf{p}),
\end{align}
where $m$ is defined in (\ref{eq:rungM}). Since $\Gamma_{\bff{p}}$ remains unchanged, the structure of the integral equation in (\ref{eq:integralEq}) is unchanged as well.  The numerics follow through exactly, hence the Lyapunov exponent for the non-commutative matrix model in the large $\Theta$ limit is,
\begin{align}
	\lambda^L_{\text{matrix}} &= 0.0019 \frac{4\lambda_1^2 + \lambda_2^2}{5\beta^2 m}.
\end{align}
Compared with the Lyapunov exponent found in \cite{Stanford:2015owe} for a commutative matrix model, the non-commutative matrix model's Lyapunov exponent, at large $\Theta$, is smaller by a factor of $5/16$ (when the comparison is done with $\lambda_1=\lambda_2$).

\section{Discussion and conclusion}\label{sec:Conclusion}
In this section, we discuss a few salient features of our computation and their implications, and conclude by mentioning some future directions.

\paragraph{Lyapunov exponent is $1/N$ suppressed in the vector case:} Unlike in the matrix case, the Lyapunov exponent is $1/N$-suppressed in the vector model, so that in the strict large-$N$ limit, the system is no longer chaotic. To explain this fact, we note that the vector model becomes weakly coupled at large $N$, unlike the matrix model. One way to see this is by noting that the anomalous dimension of conserved currents in the vector model is $1/N$-suppressed \cite{Giombi:2016ejx}. This fact was also noted in \cite{Chowdhury:2017jzb} for the vector model at criticality.

\paragraph{The large-$\Theta$ limit doesn't change $\lambda_{L}$ by much:} We found that - for both the vector case and the matrix case - $\lambda_{L}$ in the large $\Theta$ limit is of the same order of magnitude, although somewhat smaller, than in the commutative limit. This is counter-intuitive, since one might naively expect noncommutativity and nonlocality to enhance chaos.

The finding above suggests that there is still a fair amount of locality which persists in the theory even at large $\Theta$. That such locality remains in the theory at large $\Theta$ can be seen by analyzing the free energy at 2-loop level.  The free energy receives contributions from the planar figure-8 vacuum diagram and the non-planar figure-8 vacuum diagram (see Appendix \ref{app:freeEnergy}). In the large $\Theta$ limit, the non-planar figure-8 vacuum diagram vanishes, leaving us with the planar diagram and a slightly smaller combinatoric factor. This means that the free energy still scales with the temperature $T$ as $T^{4}$, a behavior consistent with a local hot gas in 3+1 dimensions.

As mentioned above, we might have expected that because of the UV-IR connection, the Lyapunov exponent would have been affected.  However, as noted by Stanford \cite{Stanford:2015owe}, the Lyapunov exponent goes like $1/m$, which means it is IR-sensitive. Note that in the limit of zero mass, there is still an infrared cutoff present proportional to $\sqrt{\lambda} ~T$. In other words, the Lyapunov exponent's largest contribution comes from low momentum modes.

The key point is that these low-momentum modes behave like quasi-local excitations. This is because the quanta in a non-commutative field theory can be thought of as dipoles moving in a magnetic field, with the transversal size of the dipole proportional to its momentum \cite{Bigatti:1999iz} -- a property of the theory known as the UV/IR connection. It follows that low-momentum modes also have small transversal size, and hence their non-locality is negligible. Perhaps this is the reason why the Lyapunov exponent is relatively insensitive to varying $\Theta$.

\paragraph{Comparison with holography:} In holography, the $\Theta$-dependence of the Lyapunov exponent has been studied in \cite{Fischler:2018kwt}, where it was found that the Lyapunov exponent is independent of $\Theta$. The holographic case is similar to the matrix case considered in our paper, since all the fields in N=4 super Yang-Mills are in the adjoint representation. An important distinction is that the holographic case only has the analog of the coupling $\mathrm{Tr}{(\Phi * \Phi * \Phi * \Phi)}$, and does not have the ``mixed-up matrix multiplication coupling.''

For these reasons, it is reasonable to expect that the Lyapunov exponent is independent of $\Theta$ at leading order in $1/N$. To see this, we first note that the only rung shapes which contribute to leading order in $1/N$ are the planar rung, the ``4-hump'' rung, and the ``2-hump'' rung (the first two are in the first diagram class of Table \ref{table:ladders}, the third one is the sixth diagram class of Table \ref{table:ladders}). These rung shapes are precisely the ones that are independent of $\Theta$ ! Hence, the sum of ladder diagrams, and consequently the Lyapunov exponent, is independent of $\Theta$, which is in agreement with the findings of \cite{Fischler:2018kwt}.

\paragraph{The relative factor of $3$:} In all the four cases, the commutative vector model, the non-commutative vector model at large $\Theta$, the commutative matrix model, and the non-commutative matrix model at large $\Theta$, the integral equation has a relative factor of $3$ between the two terms on the right-hand side,
\begin{equation}
    -i\omega f(\omega,\textbf{p}) = \int \frac{d^{3}k}{(2\pi)^3} m(\textbf{k},\textbf{p}) \bigg( f(\omega,\textbf{k}) - \frac{\sinh{(\beta E_{\textbf{p}}/2)}}{3 \sinh{(\beta E_{\textbf{k}}/2)}} f(\omega,\textbf{p}) \bigg).
\end{equation}
As noted in \cite{Stanford:2015owe}, the integral equation has the form of a Boltzmann equation, and the relative factor of $3$ can be interpreted as meaning that each collision involving an infected particle results in the loss of one and the creation of three particles. So, if that relative factor were smaller (larger) than $3$, the system would be less (more) chaotic. The fact that this relative factor is universal between the cases considered means that the physics is roughly the same.

\paragraph{Exact $\Theta$ analysis:} In the large $\Theta$ case, we have not included in our analysis subleading contributions coming from highly oscillatory diagrams. It would be interesting to explore such contributions in future work. Even though they are subleading contributions, they may contain interesting information. For example, at the level of the free energy of the field theory, such subleading contributions have been argued to look like lower-dimensional systems \cite{Fischler:2000fv, Fischler:2000bp}.

\section{Acknowledgments}
The authors would like to thank Josiah Couch, Douglas Stanford and Brian Swingle for useful discussions. The work of WF and TG is supported by the National Science Foundation under Grant Number PHY-1914679. PN acknowledges support from
Israel Science Foundation grant 447/17 for the work in sections 3 and from U.S. National Science Foundation grant PHY-1820734 for the work in section 4.

\appendix

\section{Double Ladders}\label{app:DoubleLadders}
Here we will outline the arguments which allow the double ladders to be dropped from the perturbation expansion in the vector model.  This series of diagrams is shown below along with the standard single ladder series.

\begin{equation*}
	\begin{tikzpicture}[baseline=(current bounding box.center)]
		\begin{feynman}
			\vertex (ui);
			\vertex [below=1.5cm of ui] (li);
			\vertex [right=1cm of ui, dot] (a) {};
			\vertex [right=1cm of li, dot] (b) {};
			\vertex [right=1.8cm of a, dot] (c) {};
			\vertex [right=1.8cm of b, dot] (d) {};
			\vertex [right=1cm of c] (uf);
			\vertex [right=1cm of d] (lf);
			
			\diagram* [horizontal=ui to a] {
				(ui) [dot] -- (a),
				(a) -- [out = 20, in = 160, max distance = 1cm] (c),
				(a) -- [out = -20, in = -160, max distance = 1cm] (c) -- (uf),
				(li) -- (b),
				(b) -- [out = 20, in = 160, max distance = 1cm] (d),
				(b) -- [out = -20, in = -160, max distance = 1cm] (d) -- (lf),
				(a) -- (b),
				(c) -- (d),
				
				ui--[opacity=0] uf,
				lf--[opacity=0] lf,
			};
		\end{feynman}
	\end{tikzpicture}
	+
	\begin{tikzpicture}[baseline=(current bounding box.center)]
		\begin{feynman}
			\vertex (ui);
			\vertex [below=1.5cm of ui] (li);
			\vertex [right=1cm of ui, dot] (a) {};
			\vertex [right=1cm of li, dot] (b) {};
			\vertex [right=1.8cm of a, dot] (c) {};
			\vertex [above right = 0.2cm and 0.9 of a, dot] (e) {};
			\vertex [above right = 0.2cm and 0.9 of b, dot] (f) {};
			\vertex [right=1.8cm of b, dot] (d) {};
			\vertex [right=1cm of c] (uf);
			\vertex [right=1cm of d] (lf);
			
			\diagram* [horizontal=ui to a] {
				(ui) [dot] -- (a),
				(a) -- [out = 20, in = 180, max distance = 1cm] (e),
				(e) -- [out = 0, in = 160, max distance = 1cm] (c),
				(a) -- [out = -20, in = -160, max distance = 1cm] (c) -- (uf),
				(li) -- (b),
				(b) -- [out = 20, in = 180, max distance = 1cm] (f),
				(f) -- [out = 0, in = 160, max distance = 1cm] (d),
				(b) -- [out = -20, in = -160, max distance = 1cm] (d) -- (lf),
				(a) -- (b),
				(c) -- (d),
				
				(e) -- [out = -45, in = 45, max distance =1cm] (f),
				(e) -- [out = -135, in = 135, max distance =1cm] (f),
				
				ui--[opacity=0] uf,
				lf--[opacity=0] lf,
			};
		\end{feynman}
	\end{tikzpicture}
	+
	\begin{tikzpicture}[baseline=(current bounding box.center)]
		\begin{feynman}
			\vertex (ui);
			\vertex [below=1.5cm of ui] (li);
			\vertex [right=1cm of ui, dot] (a) {};
			\vertex [right=1cm of li, dot] (b) {};
			\vertex [right=1.8cm of a, dot] (c) {};
			\vertex [right=1.8cm of b, dot] (d) {};
			
			\vertex [above right = 0.2cm and 0.9 of a, dot] (e) {};
			\vertex [above right = 0.2cm and 0.9 of b, dot] (f) {};
			
			\vertex [below left = 0.18cm and 0.45 of c, dot] (g) {};
			\vertex [below left = 0.18cm and 0.45 of d, dot] (h) {};
			
			\vertex [right=1cm of c] (uf);
			\vertex [right=1cm of d] (lf);
			
			\diagram* [horizontal=ui to a] {
				(ui) [dot] -- (a),
				
				(a) -- [out = 20, in = 180, max distance = 1cm] (e),
				(e) -- [out = 0, in = 160, max distance = 1cm] (c),
				
				(a) -- [out = -20, in = -150, max distance = 1cm] (g),
				(g) -- [out = 30, in = -160, max distance = 1cm] (c),				
				
				(c) -- (uf),
				
				(li) -- (b),
				
				(b) -- [out = 20, in = 180, max distance = 1cm] (f),
				(f) -- [out = 0, in = 160, max distance = 1cm] (d),
				
				(b) -- [out = -20, in = -150, max distance = 1cm] (h),
				(h) -- [out = 30, in = -160, max distance = 1cm] (d),
				
				(d) -- (lf),
				
				(a) -- (b),
				(c) -- (d),
				
				(e) -- [out = -45, in = 45, max distance =1cm] (f),
				(e) -- [out = -135, in = 135, max distance =1cm] (f),
				
				(g) -- [out = -45, in = 45, max distance =1cm] (h),
				(g) -- [out = -135, in = 135, max distance =1cm] (h),
				
				ui--[opacity=0] uf,
				lf--[opacity=0] lf,
			};
		\end{feynman}
	\end{tikzpicture}
	+ \cdots
\end{equation*}
\begin{equation*}
	\begin{tikzpicture}[baseline=(current bounding box.center)]
		\begin{feynman}
			\vertex (ui);
			\vertex [below=1.5cm of ui] (li);
			\vertex [right=3cm of ui] (uf);
			\vertex [right=3cm of li] (lf);
			
			\diagram* [horizontal=ui to a] {
				(ui) -- (uf),
				(li) -- (lf),
				
				ui--[opacity=0] uf,
				lf--[opacity=0] lf,
			};
		\end{feynman}
	\end{tikzpicture}
	+
	\begin{tikzpicture}[baseline=(current bounding box.center)]
		\begin{feynman}
			\vertex (ui);
			\vertex [below=1.5cm of ui] (li);
			\vertex [right=1.5cm of ui, dot] (a) {};
			\vertex [right=1.5cm of li, dot] (b) {};
			
			\vertex [right=1.5cm of a] (uf);
			\vertex [right=1.5cm of b] (lf);
			
			\diagram* [horizontal=ui to a] {
				(ui) -- (a) -- (uf),
				(li) -- (b) -- (lf),
				(a) -- [out = -135, in=135, max distance =1cm] (b),
				(a) -- [out = -45, in=45, max distance =1cm] (b),
				
				ui--[opacity=0] uf,
				lf--[opacity=0] lf,
			};
		\end{feynman}
	\end{tikzpicture}
	+
	\begin{tikzpicture}[baseline=(current bounding box.center)]
		\begin{feynman}
			\vertex (ui);
			\vertex [below=1.5cm of ui] (li);
			\vertex [right=1cm of ui, dot] (a) {};
			\vertex [right=1cm of li, dot] (b) {};
			
			\vertex [right=1cm of a, dot] (c) {};
			\vertex [right=1cm of b, dot] (d) {};
			
			\vertex [right=1cm of c] (uf);
			\vertex [right=1cm of d] (lf);
			
			\diagram* [horizontal=ui to a] {
				(ui) -- (a) -- (c) -- (uf),
				(li) -- (b) -- (d) -- (lf),
				(a) -- [out = -135, in=135, max distance =1cm] (b),
				(a) -- [out = -45, in=45, max distance =1cm] (b),
				
				(c) -- [out = -135, in=135, max distance =1cm] (d),
				(c) -- [out = -45, in=45, max distance =1cm] (d),
				
				ui--[opacity=0] uf,
				lf--[opacity=0] lf,
			};
		\end{feynman}
	\end{tikzpicture}
	+\cdots
\end{equation*}

The progression of $N$ scaling in these diagrams goes like $N^0, 1/N, 1/N^2,\dots$. However, if we look at the progression of $N$ scalings in the single ladder case, we have, $N, N^0, 1/N, \dots$.  Hence, the overall growth of the OTOC will be dominated by the single ladders. More concretely,
\begin{align}
	C_{\text{single}} &\sim \frac{1}{N} e^{\lambda_S t/N}\\
	C_{\text{double}} &\sim \frac{1}{N^2} e^{\lambda_D t/N}.
\end{align}
It seems likely that the exponential growth of the double-ladders is the same as the one computed in this paper for the single-ladders, because the same rungs are involved.  It should be noted that if, in fact $\lambda_S < \lambda_D$, then the argument for dropping the double ladders would fail since the double ladders would then dominate in the late $t$ limit.  We conjecture that $\lambda_D$ is at most equal to $\lambda_S$, however, this deserves a more careful study.

\section{Diagram Classes}\label{app:DiagramClasses}
Table \ref{table:ladders} makes reference to the number of diagrams in a particular class -- here we will explain exactly what those classes are.  In particular, a diagram class is described by its vertex factor, and hence all diagrams with the same vertex factor can be grouped into one class.

\paragraph{Class 1:} The first class includes two diagrams that we call the standard diagram, and the 4-hump diagram.  The standard diagram is the one shown in the table.  The 4-hump diagram is given by

\feynmandiagram [small, horizontal=i1 to a] {
	f1 -- b[dot] -- f2,
	i1 -- a[dot] -- i2,
	b -- [out = -45, in = 45, min distance = 1.5cm] a,
	b -- [out = -135, in = 135, min distance = 1.5cm] a,
	i1--[opacity=0]f1,
	i2--[opacity=0]f2,
};

\paragraph{Class 2:} The second class is similar to the first class in that there is one twisted standard diagram and one twisted 4-hump diagram.  The twisted 4-hump diagram is similar to the 4-hump diagram pictured above, however its loop lines are crossed like the diagram show in the table.

\paragraph{Class 3:} The third class contains a total of eight diagrams.  Four of these are similar to the 1-hump diagram show in the table except that the hump crosses over different external legs for a total of four 1-hump diagrams.  There are also four permutations of the so-called 3-hump diagram shown below.

\feynmandiagram [small, horizontal=i1 to a] {
	f1 -- b[dot] -- f2,
	i1 -- a[dot] -- i2,
	b -- [out = 45, in = 45, min distance = 1.5cm] a,
	b -- [out = -135, in = 135, min distance = 1.5cm] a,
	i1--[opacity=0]f1,
	i2--[opacity=0]f2,
};

\paragraph{Classes 4 and 5:} The fourth and fifth classes of diagrams contains two heart diagrams and two twisted heart diagrams, respectively. They are similar to the ones shown in the table except that the two humps are on the bottom.

\paragraph{Classes 6 and 7:} These are self explanatory since there is only one diagram in each class, and no other diagrams can give these vertices.

\section{Free Energy in $\phi^4$ theory} \label{app:freeEnergy}
Here we will compute the two-loop correction to the free energy in non-commutative $\lambda \phi^4$ theory in the large $\Theta$ limit.  We will see how the non-planar contributions vanish in this limit, and the end result is to modify the planar contributions so that the combinatoric factor is distinct from the commutative case.  

The first non-trivial correction to the free energy appears at order $\lambda$, and is given diagrammatically by the two loop diagrams,
\begin{equation}
	\feynmandiagram [baseline=(a.base), horizontal = a, layered layout] {
		a [dot] 
		-- [out = 135, in = 45, loop, min distance = 2cm] a
		-- [out = -135, in = -45, loop, min distance = 2cm] a
	};
\hspace{.2\linewidth}
	\feynmandiagram [baseline=(a.base), horizontal = a, layered layout] {
		a [dot] 
		-- [out = 90, in = -45, loop, min distance = 2.5cm] a
		-- [out = 20, in = -100, loop, min distance = 3.5cm] a
	};
\end{equation}
The planar diagram on the left is given by,
\begin{align} \label{eq:vacDiagramPlanar}
	-\frac{\lambda T^2}{2} \sum_{n=-\infty}^{\infty} \sum_{l=-\infty}^{\infty} \int\frac{d^3\bff{p}}{(2\pi)^3} \int\frac{d^3\bff{k}}{(2\pi)^3} \frac{1}{\left(\frac{4\pi^2n^2}{\beta^2} + \bff{p}^2 + m^2\right) \left(\frac{4\pi^2l^2}{\beta^2} + \bff{k}^2 + m^2\right)},
\end{align}
while the non-planar diagram on the right is given by,
\begin{align}
	-\frac{\lambda T^2}{4} \sum_{n=-\infty}^{\infty} \sum_{l=-\infty}^{\infty} \int\frac{d^3\bff{p}}{(2\pi)^3} \int\frac{d^3\bff{k}}{(2\pi)^3} \frac{e^{i \Theta\cdot p\cdot k}}{\left(\frac{4\pi^2n^2}{\beta^2} + \bff{p}^2 + m^2\right) \left(\frac{4\pi^2l^2}{\beta^2} + \bff{k}^2 + m^2\right)}.
\end{align}
Now, focusing on the second diagram, and setting $\Theta^{12} = -\Theta^{21} = \Theta$, with all other components $0$, we find after doing the $\bff{k}$ integral \cite{Fischler:2000fv},
\begin{align}
	-\frac{\lambda T^2}{4} \sum_{n,l} \int\frac{d^3\bff{p}}{(2\pi)^3} \frac{1}{\left(\frac{4\pi^2n^2}{\beta^2} + \bff{p}^2\right) \left(4\pi^2 \left(l^2\beta^2 + \abs{\Theta \cdot p}^2\right) + \Lambda^{-2}\right)},
\end{align}
where $\abs{\Theta\cdot p} = \Theta\sqrt{p_1^2 + p_2^2}$, and $\Lambda$ is a UV cutoff.  The mass is being neglected here since we are still working in the $\beta m \ll 1$ limit.  Now performing the sums over $n$ and $l$, we find,
\begin{align}
	-\frac{\lambda T^2}{4} \int\frac{d^3\bff{p}}{(2\pi)^3} \frac{1 + 2 n_\beta\left(\abs{\bff{p}}\right)}{2 \abs{\bff{p}}} \frac{1 + 2 n_{1/\beta} \left(2\pi \abs{\Theta\cdot p}\right)}{4\pi \abs{\Theta\cdot p}},
\end{align}
where $n_\beta(x) = \frac{1}{e^{\beta x} - 1}$ is a Bose distribution at temperature $T = 1/\beta$.

It is easy to see that in the large $\Theta$ limit, the entire expression goes like $\sim 1/\Theta$, and hence vanishes.  Thus, the only contribution to the free energy at order $\lambda$ comes from the planar diagram (\ref{eq:vacDiagramPlanar}).  This is exactly the same as the commutative case except that the factor in front of the integral is $3 \lambda T^2/4$ in the commutative case.

So we see again that in the large $\Theta$ limit, the only diagrams that survive are the same as the commutative case with the exception that the combinatoric factors are decreased by some factor.

\newpage
\pagestyle{empty}
\begin{table}[t!]
	\begin{threeparttable}
		\centering
		\renewcommand{\arraystretch}{2}
		\caption{Ladder diagrams}
		\label{table:ladders}
		\begin{tabular}{ || c | c | c | c || }
			\hline
			Diagram Class & Diagrams/Class & Prefactor & Vertex Factor\\
			\hline
			\hline
			\feynmandiagram [small, horizontal=i1 to a] {
				f1 -- b[dot] -- f2,
				i1 -- a[dot] -- i2,
				b -- [out = 45, in = 320, max distance = 1cm] a,
				b -- [out = 135, in = 225, max distance = 1cm] a,
				i1--[opacity=0]f1,
				i2--[opacity=0]f2,
			};& 2 & $\frac12\lambda_1^2 + \lambda_2^2$ & $e^{\frac{i}{2} \Theta \cdot \omega \cdot (p - p')}$\\
			\hline
			\feynmandiagram [small, horizontal=i1 to a] {
				f1 -- b[dot] -- f2,
				i1 -- a[dot] -- i2,
				b -- [out = 45, in = -135, max distance = 1cm] a,
				b -- [out = 135, in = -45, max distance = 1cm] a,
				i1--[opacity=0]f1,
				i2--[opacity=0]f2,
			};& 2 & $\frac14 \lambda_1^2 + \lambda_1\lambda_2$ & $e^{\frac{i}{2} \Theta \cdot (\omega \cdot p - \omega \cdot p' -2p\cdot \ell +2p'\cdot \ell)}$\\
			\hline
			\feynmandiagram [small, horizontal=i1 to a] {
				f1 -- b[dot] -- f2,
				i1 -- a[dot] -- i2,
				b -- [out = 45, in = 45, min distance = 1.5cm] a,
				b -- [out = 135, in = 225, min distance = 1cm] a,
				i1--[opacity=0]f1,
				i2--[opacity=0]f2,
			};& 8 & $\frac14 \lambda_1^2 + \lambda_1\lambda_2$ & $e^{\frac{i}{2} \Theta \cdot (\omega \cdot p - \omega \cdot p' - 2p \cdot \ell)}$\\
			\hline
			\feynmandiagram [small, horizontal=i1 to a] {
				f1 -- b[dot] -- f2,
				i1 -- a[dot] -- i2,
				b -- [out = 45, in = 45, min distance = 1.5cm] a,
				b -- [out = 135, in = 135, min distance = 1.5cm] a,
				i1--[opacity=0]f1,
				i2--[opacity=0]f2,
			};& 2 & $\frac12\lambda_1^2 + \lambda_2^2$ & $e^{\frac{i}{2} \Theta \cdot (\omega \cdot p - \omega \cdot p' + 2p \cdot p' - 2p \cdot \ell + 2p' \cdot \ell)}$\\
			\hline
			\feynmandiagram [small, horizontal=i1 to a] {
				f1 -- b[dot] -- f2,
				i1 -- a[dot] -- i2,
				b -- [out = 45, in = 135, min distance = 1cm] a,
				b -- [out = 135, in = 45, min distance = 1cm] a,
				i1--[opacity=0]f1,
				i2--[opacity=0]f2,
			};& 2 & $\frac14\lambda_1^2 + \lambda_1\lambda_2$ & $e^{\frac{i}{2} \Theta \cdot (\omega \cdot p - \omega \cdot p' + 2p \cdot p')}$\\
			\hline
			\feynmandiagram [small, horizontal=i1 to a] {
				f1 -- b[dot] -- f2,
				i1 -- a[dot] -- i2,
				b -- [out = -45, in = 45, min distance = 1.5cm] a,
				b -- [out = 135, in = -135, min distance = 1cm] a,
				i1--[opacity=0]f1,
				i2--[opacity=0]f2,
			};& 1 & $\frac12 \lambda_1^2 + \lambda_2^2$ & $e^{\frac{i}{2} \Theta \cdot \omega \cdot (p - p' - 2\ell)}$\\
			\hline
			\feynmandiagram [small, horizontal=i1 to a] {
				f1 -- b[dot] -- f2,
				i1 -- a[dot] -- i2,
				b -- [out = 45, in = 45, min distance = 1.5cm] a,
				b -- [out = -135, in = -135, min distance = 1.5cm] a,
				i1--[opacity=0]f1,
				i2--[opacity=0]f2,
			};& 1 & $\frac12 \lambda_1^2 + \lambda_2^2$ & $e^{\frac{i}{2} \Theta \cdot (\omega \cdot p - \omega \cdot p' - 2\omega \cdot \ell + 2p \cdot p' + 2p \cdot \ell + 2p' \cdot \ell)}$\\
			\hline
		\end{tabular}
		\begin{tablenotes}
			\small
			\item \emph{The column, Diagrams/Class, indicates how many diagrams have the same vertex factor and prefactor as show in the third and fourth columns. Hence, the overall prefactor should be the product of columns two and three.}
		\end{tablenotes}
	\end{threeparttable}
\end{table}

\begin{center}
	\begin{table}[t!]
		\begin{threeparttable}
			\renewcommand{\arraystretch}{2}
			\caption{2-loop melon diagrams}
			\label{table:melons}
			\centering
			\begin{tabular}{ || c | c | c || }
				\hline
				Diagram Class & Prefactor & Vertex Factor\\
				\hline
				\hline
				\feynmandiagram [horizontal=i1 to a, layered layout] {
					i1 -- a[dot],
					b[dot] -- i2,
					a -- [out = 65, in = 115, max distance = 1cm] b,
					a -- [out = 0, in = 180, max distance = 1cm] b,
					a -- [out = -65, in = -115, max distance = 1cm] b,			
				};& $\frac12\lambda_1^2 + \lambda_2^2$ & $1$\\
				\hline
				\feynmandiagram [horizontal=i1 to a, layered layout] {
					i1 -- a[dot],
					b[dot] -- i2,
					a -- [out = 65, in = -115, max distance = 1cm] b,
					a -- [out = 0, in = 180, max distance = 1cm] b,
					a -- [out = -65, in = 115, max distance = 1cm] b,	
				};& $\frac12 \lambda_1^2 + \lambda_2^2$ & $e^{-\frac{i}{2} \Theta \cdot (2pp' + 2pp'' + 2p''p')}$\\
				\hline
				\feynmandiagram [horizontal=i1 to a, layered layout] {
					i1 -- a[dot],
					b[dot] -- i2,
					a -- [out = 65, in = 180, max distance = 1cm] b,
					a -- [out = 0, in = 115, max distance = 1cm] b,
					a -- [out = -65, in = -115, max distance = 1cm] b,	
				};& $\frac14 \lambda_1^2 + \lambda_1\lambda_2$ & $e^{-\frac{i}{2} \Theta \cdot (2p''p')}$\\
				\hline
				\feynmandiagram [horizontal=i1 to a, layered layout] {
					i1 -- a[dot],
					b[dot] -- i2,
					a -- [out = 65, in = 115, max distance = 1cm] b,
					a -- [out = 0, in = -115, max distance = 1cm] b,
					a -- [out = -65, in = 180, max distance = 1cm] b,	
				};& $\frac14 \lambda_1^2 + \lambda_1\lambda_2$ & $e^{-\frac{i}{2} \Theta \cdot (2p'p'')}$\\
				\hline
				\feynmandiagram [horizontal=i1 to a, layered layout] {
					i1 -- a[dot],
					b[dot] -- i2,
					a -- [out = 55, in = 180, max distance = 1cm] b,
					a -- [out = 0, in = -125, max distance = 1cm] b,
					a -- [out = -65, in = 115, max distance = 1cm] b,	
				};& $\frac14 \lambda_1^2 + \lambda_1\lambda_2$ & $e^{-\frac{i}{2} \Theta \cdot (2p'p)}$\\
				\hline
				\feynmandiagram [horizontal=i1 to a, layered layout] {
					i1 -- a[dot],
					b[dot] -- i2,
					a -- [out = 55, in = -115, max distance = 1cm] b,
					a -- [out = 0, in = 115, max distance = 1cm] b,
					a -- [out = -65, in = 180, max distance = 1cm] b,	
				};& $\frac14 \lambda_1^2 + \lambda_1\lambda_2$ & $e^{-\frac{i}{2} \Theta \cdot (2pp')}$\\
				\hline
			\end{tabular}
			\begin{tablenotes}
				\small
				\item \emph{The prefactor column indicates the coupling and combinatoric factors for each diagram.}
			\end{tablenotes}
		\end{threeparttable}
	\end{table}
\end{center}

\begin{thebibliography}{99}

\bibitem{KitaevTalk}
A.~Kitaev,
``Hidden correlations in the Hawking radiation and thermal noise,''
talk given at KITP, 2015. URL: https://online.kitp.ucsb.edu/online/joint98/kitaev/

\bibitem{Shenker:2013pqa}
S.~H.~Shenker and D.~Stanford,
``Black holes and the butterfly effect,''
JHEP \textbf{03}, 067 (2014)
[arXiv:1306.0622 [hep-th]].

\bibitem{Shenker:2014cwa}
S.~H.~Shenker and D.~Stanford,
``Stringy effects in scrambling,''
JHEP \textbf{05}, 132 (2015)
doi:10.1007/JHEP05(2015)132
[arXiv:1412.6087 [hep-th]].

\bibitem{Maldacena:2015waa}
J.~Maldacena, S.~H.~Shenker and D.~Stanford,
``A bound on chaos,''
JHEP \textbf{08}, 106 (2016)
doi:10.1007/JHEP08(2016)106
[arXiv:1503.01409 [hep-th]].

\bibitem{Maldacena:2016hyu}
J.~Maldacena and D.~Stanford,
``Remarks on the Sachdev-Ye-Kitaev model,''
Phys. Rev. D \textbf{94}, no.10, 106002 (2016)
doi:10.1103/PhysRevD.94.106002
[arXiv:1604.07818 [hep-th]].

\bibitem{Maldacena:2017axo}
J.~Maldacena, D.~Stanford and Z.~Yang,
``Diving into traversable wormholes,''
Fortsch. Phys. \textbf{65}, no.5, 1700034 (2017)
doi:10.1002/prop.201700034
[arXiv:1704.05333 [hep-th]].

\bibitem{deBoer:2017xdk}
J.~de Boer, E.~Llabr\'es, J.~F.~Pedraza and D.~Vegh,
``Chaotic strings in AdS/CFT,''
Phys. Rev. Lett. \textbf{120}, no.20, 201604 (2018)
doi:10.1103/PhysRevLett.120.201604
[arXiv:1709.01052 [hep-th]].

\bibitem{Fischler:2018kwt}
W.~Fischler, V.~Jahnke and J.~F.~Pedraza,
``Chaos and entanglement spreading in a non-commutative gauge theory,''
JHEP \textbf{11}, 072 (2018)
[erratum: JHEP \textbf{02}, 149 (2021)]
doi:10.1007/JHEP11(2018)072
[arXiv:1808.10050 [hep-th]].

\bibitem{Couch:2019zni}
J.~Couch, S.~Eccles, P.~Nguyen, B.~Swingle and S.~Xu,
``Speed of quantum information spreading in chaotic systems,''
Phys. Rev. B \textbf{102}, no.4, 045114 (2020)
doi:10.1103/PhysRevB.102.045114
[arXiv:1908.06993 [cond-mat.stat-mech]].

\bibitem{Eccles:2021zum}
S.~Eccles, W.~Fischler, T.~Guglielmo, J.~F.~Pedraza and S.~Racz,
``Speeding up the spread of quantum information in chaotic systems,''
[arXiv:2108.12688 [hep-th]].

\bibitem{Romero-Bermudez:2019vej}
A.~Romero-Berm\'udez, K.~Schalm and V.~Scopelliti,
``Regularization dependence of the OTOC. Which Lyapunov spectrum is the physical one?,''
JHEP \textbf{07}, 107 (2019)
doi:10.1007/JHEP07(2019)107
[arXiv:1903.09595 [hep-th]].

\bibitem{Liao:2018uxa}
Y.~Liao and V.~Galitski,
``Nonlinear sigma model approach to many-body quantum chaos: Regularized and unregularized out-of-time-ordered correlators,''
Phys. Rev. B \textbf{98}, no.20, 205124 (2018)
doi:10.1103/PhysRevB.98.205124
[arXiv:1807.09799 [cond-mat.dis-nn]].

\bibitem{Kundu:2021qcx}
S.~Kundu,
``Subleading Bounds on Chaos,''
[arXiv:2109.03826 [hep-th]].

\bibitem{Kundu:2021mex}
S.~Kundu,
``Extremal Chaos,''
[arXiv:2109.08693 [hep-th]].

\bibitem{Edalati:2012jj}
M.~Edalati, W.~Fischler, J.~F.~Pedraza and W.~Tangarife Garcia,
``Fast Scramblers and Non-commutative Gauge Theories,''
JHEP \textbf{07}, 043 (2012)
doi:10.1007/JHEP07(2012)043
[arXiv:1204.5748 [hep-th]].

\bibitem{Couch:2017yil}
J.~Couch, S.~Eccles, W.~Fischler and M.~L.~Xiao,
``Holographic complexity and noncommutative gauge theory,''
JHEP \textbf{03}, 108 (2018)
doi:10.1007/JHEP03(2018)108
[arXiv:1710.07833 [hep-th]].


\bibitem{Stanford:2015owe}
D.~Stanford,
``Many-body chaos at weak coupling,''
JHEP \textbf{10}, 009 (2016)
doi:10.1007/JHEP10(2016)009
[arXiv:1512.07687 [hep-th]].

\bibitem{Minwalla:1999px}
S.~Minwalla, M.~Van Raamsdonk and N.~Seiberg,
``Noncommutative perturbative dynamics,''
JHEP \textbf{02}, 020 (2000)
doi:10.1088/1126-6708/2000/02/020
[arXiv:hep-th/9912072 [hep-th]].

\bibitem{Giombi:2016ejx}
S.~Giombi,
``Higher Spin \textemdash{} CFT Duality,''
doi:10.1142/9789813149441\_0003
[arXiv:1607.02967 [hep-th]].

\bibitem{Chowdhury:2017jzb}
D.~Chowdhury and B.~Swingle,
``Onset of many-body chaos in the $O(N)$ model,''
Phys. Rev. D \textbf{96}, no.6, 065005 (2017)
doi:10.1103/PhysRevD.96.065005
[arXiv:1703.02545 [cond-mat.str-el]].

\bibitem{Fischler:2000fv}
W.~Fischler, E.~Gorbatov, A.~Kashani-Poor, S.~Paban, P.~Pouliot and J.~Gomis,
``Evidence for winding states in noncommutative quantum field theory,''
JHEP \textbf{05}, 024 (2000)
doi:10.1088/1126-6708/2000/05/024
[arXiv:hep-th/0002067 [hep-th]].

\bibitem{Fischler:2000bp}
W.~Fischler, E.~Gorbatov, A.~Kashani-Poor, R.~McNees, S.~Paban and P.~Pouliot,
``The Interplay between theta and T,''
JHEP \textbf{06}, 032 (2000)
doi:10.1088/1126-6708/2000/06/032
[arXiv:hep-th/0003216 [hep-th]].

\bibitem{Bigatti:1999iz}
D.~Bigatti and L.~Susskind,
``Magnetic fields, branes and noncommutative geometry,''
Phys. Rev. D \textbf{62}, 066004 (2000)
doi:10.1103/PhysRevD.62.066004
[arXiv:hep-th/9908056 [hep-th]].

\end{thebibliography}
\end{document}